\documentclass[twocolumn]{aastex62}

\usepackage{apjfonts}

\submitjournal{ApJ}

\shorttitle{\HST\/ Astrometry of $\mu$~Cas}
\shortauthors{Bond et al.}

\def\mucas{$\mu$~Cas}

\newcommand{\Gaia}{{\it Gaia}}
\newcommand{\Hipp}{{\it Hipparcos}}        

\newcommand{\HST}{{\it HST}}

\newcommand{\Teff}{T_{\rm eff}}

\newcommand{\kms}{{\>\rm km\>s^{-1}}}

\newcommand{\Mjup}{M_{\rm Jup}}

\begin{document}

\title{{\em Hubble Space Telescope\/} Astrometry of the Metal-Poor Visual Binary $\mu$~Cassiopeiae: \\ Dynamical Masses, Helium Content, and Age\footnote{Based on observations with the NASA/ESA {\it Hubble Space Telescope\/} obtained at Space Telescope Science Institute, operated by Association of Universities for Research in Astronomy, Inc., under NASA contract NAS5-26555.} 
}

\author[0000-0003-1377-7145]{Howard E. Bond}
\affil{Department of Astronomy \& Astrophysics, Pennsylvania State University, University Park, PA 16802, USA}
\affil{Space Telescope Science Institute, 
3700 San Martin Dr.,
Baltimore, MD 21218, USA}

\author[0000-0001-5415-9189]{Gail H. Schaefer}
\affil{The CHARA Array of Georgia State University, Mount Wilson Observatory, Mount Wilson, CA 91023, USA}

\author[0000-0002-1554-5578]{Ronald L. Gilliland}
\affil{Department of Astronomy \& Astrophysics, Pennsylvania State University, University Park, PA 16802, USA}
\affil{Space Telescope Science Institute, 
3700 San Martin Dr.,
Baltimore, MD 21218, USA}

\author[0000-0003-3277-7685]{Don A. VandenBerg}
\affil{Department of Physics \& Astronomy, University of Victoria, P.O. Box 1700 STN CSC, Victoria, B.C. V8W 2Y2, Canada}

\correspondingauthor{Howard E. Bond}
\email{heb11@psu.edu}

\begin{abstract}

$\mu$~Cassiopeiae is a nearby, high-velocity, metal-poor ($\rm[Fe/H]=-0.81$) visual binary. We have used high-resolution imaging with the {\it Hubble Space Telescope\/} (\HST), obtained over nearly two decades, to determine the period (21.568~yr) and precise orbital elements. Combining these with published ground- and space-based astrometry, we determined dynamical masses for both components of \mucas: $0.7440\pm0.0122\,M_\odot$ for the G5~V primary, and $0.1728\pm0.0035\,M_\odot$ for its faint dM companion. We detect no significant perturbations in the \HST\/ astrometry due to a third body in the system. The primary aim of our program was to determine, with the aid of stellar models, the helium content and age of the metal-deficient primary star, \mucas~A\null. Although we now have a precise mass, there remain uncertainties about other parameters, including its effective temperature. Moreover, a re-examination of archival interferometric observations leads to a suspicion that the angular diameter was overestimated by a few percent. In the absolute magnitude versus color plane, \mucas~A lies slightly cooler and more luminous than the main sequence of the globular cluster 47~Tucanae; this may imply that the star has a lower helium content, and/or is older, and/or has a higher metallicity, than the cluster. Our best estimates for the helium content and age of \mucas~A { are $Y=0.255\pm0.014$ and $12.7\pm2.7$~Gyr}---making \mucas\ possibly the oldest star in the sky visible to the naked eye. Improved measurements of the absolute parallax of the system, the effective temperature of \mucas~A, and its angular diameter would provide tighter constraints. 

\end{abstract}




\section{$\mu$~Cassiopeiae: An Important Metal-Poor Visual Binary }



The nearby fifth-magnitude G5~V star $\mu$~Cassiopeiae was one of the first ``high-velocity'' stars to be recognized \citep{Campbell1901, Adams1919, Oort1926, Miczaika1940, Roman1955}. With a radial velocity (RV) of $-97\,\kms$ \citep[][and references therein]{Agati2015}, and a distance of 7.55~pc and proper motion of $3\farcs78\,\rm yr^{-1}$ \citep[both values from the \Hipp\/ mission;][]{vanLeeuwen2007}, the star has a total space motion relative to the Sun of $167\,\kms$ and can be considered a thick-disk or possibly a halo object. \citet{Johnson1953}, in their classical paper that introduced $UBV$ photometry, noted that \mucas\ lies below the main sequence in the color-absolute magnitude diagram for nearby stars with accurate distances, and that its $U-B$ index is relatively blue for its $B-V$ color. It was soon recognized that the low luminosities and ultraviolet excesses of high-velocity dwarfs are the result of low heavy-element contents. As discussed below, modern spectroscopic analyses of \mucas\ give a photospheric metal abundance of about 1/6 that of the Sun ($\rm[Fe/H]\simeq-0.8$).

Photographic positional measurements of \mucas\ over a quarter of a century at the Allegheny Observatory led to the discovery that the star is an astrometric binary, showing a perturbation due to an unseen companion with an orbital period of about 23~yr \citep{Wagman1961, Wagman1963}. An astrometric analysis of photographic plates from the Sproul Observatory, also over an interval of about a quarter century, refined the orbital period to $\sim$18.5~yr \citep{Lippincott1964}. Based on additional Sproul material, \citet{Lippincott1981} updated the period again to 21.43~yr; and then \citet{Russell1984} further revised the astrometric-perturbation period to 22.09~yr, based on Allegheny photographic material covering 45 years. A final analysis of all of the Sproul plates, now covering 55~years, gave a period of 21.40~yr \citep{HeintzCantor1994}.

The astrophysical importance of \mucas\ was emphasized by \citet[][hereafter D65]{Dennis1965}. The primordial abundance of helium, and the history of its increase over cosmic time due to stellar nucleosynthesis, are important constraints on cosmology and Galactic evolution. However, the old stellar populations that have survived to the present epoch contain primarily cool main-sequence stars and red giants, lacking helium lines in their spectra. D65 argued that, because of its binary nature, \mucas\ offers the possibility of
determining the helium content in an old, metal-poor star through an alternative approach. By measuring its dynamical mass, and thus the position of \mucas\ in the mass-luminosity plane, one can infer its interior helium content using theoretical stellar models. A note of caution, however, was issued by \citet{Faulkner1971}, who noted that a useful cosmological constraint would require very precise knowledge of the dynamical masses of the binary. \citet{Haywood1992} gave an interpolation formula for the dependence of the derived helium mass fraction, $Y$, on measured mass; it shows that the inferred value of $Y$ decreases by about 0.01 per increase in mass of $0.01\,M_\odot$. Thus a meaningful constraint on the He content requires a mass of \mucas~A known to better than $\sim$0.01--$0.02\,M_\odot$.

D65's paper inspired observers to attempt to detect the \mucas\ companion---which was, however, expected to be extremely faint and difficult. D65 predicted \mucas~B to be an M~dwarf with a visual magnitude difference relative to \mucas~A of $\sim$6 to 8~mag. The anticipated angular separation, reaching a maximum in the mid 1960's and then rapidly decreasing, was a little more than one second of arc. Almost a decade passed before the first successful resolution of the pair, using stellar interferometry, was reported by \citet{Wickes1974}; they also reference several earlier failed attempts.\footnote{An earlier visual resolution was reported in a conference abstract by \citet{Wehinger1966}, but the measurements (separation, position angle, and magnitude difference) are so discordant with subsequent findings that the detection appears to be spurious. According to \citet{Feibelman1976}, the claim was later withdrawn. Feibelman himself reported a partial resolution of the binary in photographs obtained in 1964 and 1965, but again his results are in poor agreement with the elements derived in subsequent work, including the present paper. \citet{Lippincott1981} lists in her Table~4 other attempts to resolve the binary up to the early 1980's. With some prescience, she stated ``One observation by the Space Telescope in combination with $\dots$ the astrometric orbit should give the total mass of the system, as well as the individual masses.'' But \citet{Pierce1985} riposted that the suggestion of a single space-based observation being sufficient would ``appear to be premature.''} They measured a separation of $0\farcs35$ and an optical magnitude difference of 5.5~mag. The resolution was confirmed in another interferometric observation a year later (Wickes 1975), but orbital motion had reduced the separation to only $0\farcs23$.

Accurate astrometry of this binary is close to the limit of what is possible with ground-based techniques, especially around periastron passage. Since the work in the 1970's, only a handful of additional ground-based measurements has been
published, as discussed below. \citet{Drummond1995} resolved the binary in two adaptive-optics observations obtained in 1994; based on the available data, they carried out an orbital solution and derived component masses of $0.742\pm0.059$ and $0.173\pm0.011\, M_\odot$. Their results implied a helium content for \mucas~A of $Y = 0.24\pm0.07$, an uncertainty too large for a worthwhile cosmological or astrophysical constraint. \citet{Horch2015,Horch2019} presented speckle astrometry of the system at three epochs; they obtained a total mass of $0.906\pm0.023\,M_\odot$, but did not derive individual masses.


By contrast, resolution of the system is relatively easy from space, based on images taken with the {\it Hubble Space Telescope\/} (\HST)\null.  In this paper, we report astrometry of \mucas, obtained with \HST\/ over an interval of nearly two decades. By combining the \HST\/ data with the ground-based measurements, we derive precise orbital elements for the binary, and the dynamical masses of both components. We also place limits on third bodies in the system. We then apply these results to a discussion of the helium content, age, and other properties of this important bright, old, and metal-poor star.

\section{{\em HST\/} Observations}

We began a program of \HST\/ imaging of \mucas\ in 1997, and continued it until 2016, for a total of 27 epochs. Observations from 1997 to 2007 were made with
the Wide Field Planetary Camera~2 (WFPC2) during 20 visits. The WFPC2 was
removed from the spacecraft during the astronaut servicing mission in 2009, and
replaced with the Wide Field Camera~3 (WFC3). We used the WFC3 UVIS channel  
for an additional seven visits to \mucas\ from 2010 to 2016. Our observations of \mucas\ were part of a long-term program of \HST\/ astrometry of astrophysically important visual binaries, which also included imaging of the Procyon and Sirius systems. Results for the latter two binaries have been published by \citet[][hereafter B15]{Bond2015} and \citet{Bond2018} for Procyon, and \citet[][hereafter B17]{Bond2017} for Sirius. 

Table~\ref{table:obslog} presents the \HST\/ observing log for \mucas. For the WFPC2 imaging,
knowing that the faint companion of \mucas\ is cooler than the primary star, we
selected the longest-wavelength filter available on the camera, which also had a
narrow enough bandpass to permit a well-exposed but unsaturated image of the
primary to be obtained in a short integration time. These considerations led to
the choice of the F953N bandpass, a narrow-band filter normally intended for
imaging of the [\ion{S}{3}] 9530~\AA\ nebular emission line. We placed \mucas\
near the center of the Planetary Camera (PC) CCD detector, providing a plate
scale of $\sim\!0\farcs0456\;\rm pix^{-1}$. For our initial two visits, we chose
exposure times of 0.3 and 0.5~s, at four dither positions each. We made these
exposures short enough to ensure that the primary's image would not be
saturated, which indeed proved to be the case. The magnitude difference between
A and B in this bandpass was measured to be 4.9~mag. Based on these frames, we
increased the integration time to 1.0~s for the next three visits, and obtained
15 dithered exposures per visit. These dithers used five different pointings, separated by several tenths of an arcsecond, and sampled five different pixel phases in both coordinates. Because images of the primary had a few
saturated pixels in a few of these frames, we reduced the exposure time to 0.8~s
for the next five visits, and took 15--17 dithered exposures at each epoch. For
the final ten visits with the aging WFPC2 instrument, we increased the exposure
time back to 1.0~s, obtaining 17 exposures per visit. For all exposures, we
chose a telescope orientation such that the faint companion would lie away from
the diffraction spikes and charge bleeding of the bright primary.

With the installation of the more-sensitive WFC3 camera, we had no available combination of a long-wavelength filter and short exposure time that would reliably produce unsaturated images of \mucas~A\null. In order to obtain unsaturated WFC3 exposures, the best choice was the ultraviolet F225W bandpass---which meant that the dM companion would be extremely faint and require long exposures for detection. We therefore adopted a strategy that we also used for Procyon (see B15): at each dither position, we obtained a short unsaturated exposure of the primary and then, without moving the telescope, a long exposure to detect the companion. For the first WFC3 visit, we obtained eight dithered 1.0~s exposures combined with eight 260~s exposures at the same pointings. To reduce data volume and avoid interruptions for buffer dumps, we used a $512\times512$ subarray for all of our WFC3 exposures. The WFC3 UVIS channel has two CCD detectors (plate scale $0\farcs0396\;\rm pix^{-1}$) with a small gap between them; for our first WFC3 visit we used UVIS1, but for the rest the better-characterized UVIS2. Based on results of the first WFC3 visit, we increased the short exposures to 1.5~s for the remainder of the F225W observations, and adjusted the long-exposure integration times slightly so as to use all of the available target visibility time during the \HST\/ orbit.

The red companion star is very faint in the WFC3 F225W filter: we measured a
difference of 9.9~mag relative to the primary. We continued to use this filter
and observing strategy for five subsequent visits, but it became apparent that
the astrometric precision was poorer than we had achieved in the far-red
bandpass used with WFPC2. The situation was becoming worse as the orbital
separation began to shrink rapidly and the companion was becoming embedded in
the wings of the primary's image. Thus, for our final observation, we adopted an
alternative approach, using the WFC3's version of the F953N filter. We obtained
dithered exposures with integration times of 0.5~s (hoping for an unsaturated
primary star---but not realized consistently), 2.5~s (for good unsaturated
exposures of the companion), and 200~s (for an attempt to centroid saturated
images of both stars using the diffraction spikes and features in the
wings---see below). 




\begin{deluxetable}{llccc}
\tablewidth{0 pt}
\tabletypesize{\footnotesize}
\tablecaption{\HST\/ Observing Log for \mucas
\label{table:obslog}}
\tablehead{
\colhead{UT Date} &
\colhead{Dataset\tablenotemark{a}} &
\colhead{Exposure } &
\colhead{No.} &
\colhead{Proposal} \\
\colhead{} &
\colhead{} &
\colhead{Time(s) [s]} &
\colhead{Frames\tablenotemark{b}} &
\colhead{ID\tablenotemark{c}} 
}
\startdata
\noalign{\smallskip}
\multispan5{\hfil WFPC2/PC Frames, F953N Filter\tablenotemark{d} \hfil} \\
1997 Jul 04 & U42K0201M & 0.3, 0.5 & 14 & 7497 \\
1998 Jan 02 & U42K0301R & 0.3, 0.5 & 14 & 7497 \\
1998 Jul 22 & U42K0401R & 1.0      & 15 & 7497 \\
1999 Feb 28 & U42K0901R & 1.0      & 15 & 7497 \\  
1999 Aug 04 & U59H0201R & 1.0      & 15 & 8396 \\ 
2000 Feb 01 & U59H0301R & 0.8      & 15 & 8396 \\ 
2000 Jul 15 & U67H0201R & 0.8      & 16 & 8586 \\
2001 Jan 15 & U67H0301R & 0.8      & 16 & 8586 \\
2001 Jul 30 & U6IZ0201R & 0.8      & 17 & 9227 \\
2002 Jan 17 & U6IZ0301M & 0.8      & 17 & 9227 \\
2002 Aug 05 & U8IP0201M & 1.0      & 17 & 9332 \\
2003 Feb 11 & U8IP0301M & 1.0      & 17 & 9332 \\
2003 Aug 05 & U8RM0201M & 1.0      & 17 & 9887 \\
2004 Jan 29 & U8RM0301M & 1.0      & 17 & 9887 \\
2004 Aug 08 & U9290201M & 1.0      & 17 & 10112 \\
2005 Jan 15 & U9290301M & 1.0      & 17 & 10112 \\
2005 Aug 13 & U9D30201M & 1.0      & 17 & 10481 \\
2006 Jan 30 & U9D30301M & 1.0      & 17 & 10481 \\
2006 Sep 26 & U9O50201M & 1.0      & 17 & 10914 \\
2007 Oct 17 & UA0P0201M & 1.0      & 17 & 11296 \\
\noalign{\smallskip}
\multispan5{\hfil WFC3/UVIS Frames, F225W Filter \hfil} \\
2010 Jan 09 & IB7J02010 & 1.0, 260 & 16 & 11786 \\
2010 Dec 03 & IBK702010 & 1.5, 265 & 16 & 12296 \\
2011 Dec 05 & IBTI02010 & 1.5, 265 & 16 & 12673 \\
2012 Dec 02 & IC1K02010 & 1.5, 236 & 16 & 13062 \\
2013 Oct 25 & ICA102010 & 1.5, 259 & 16 & 13468 \\
2015 Jan 06 & ICJX02010 & 1.5, 255 & 16 & 13876 \\
\noalign{\smallskip}
\multispan5{\hfil WFC3/UVIS Frames, F953N Filter \hfil} \\
2016 Jul 11 & ICVD02010 & 0.5, 2.5, 200 & 24 & 14342 \\
\enddata
\tablenotetext{a}{Dataset identifier for first observation made at each visit. 
}
\tablenotetext{b}{Total number of individual frames obtained during each visit.
}
\tablenotetext{c}{\HST\/ proposal identification number. H.E.B. was PI for all of these programs.
}
\tablenotetext{d}{0.11~s exposures were also taken in F467M and F547M on 1997 Jul 04 and 1998 Jan 02, in an attempt to determine the color of the companion; however, it was not detected in these frames.
}
\end{deluxetable}


%
%

\begin{figure*}[ht]
\centering
\includegraphics[width=5.5in]{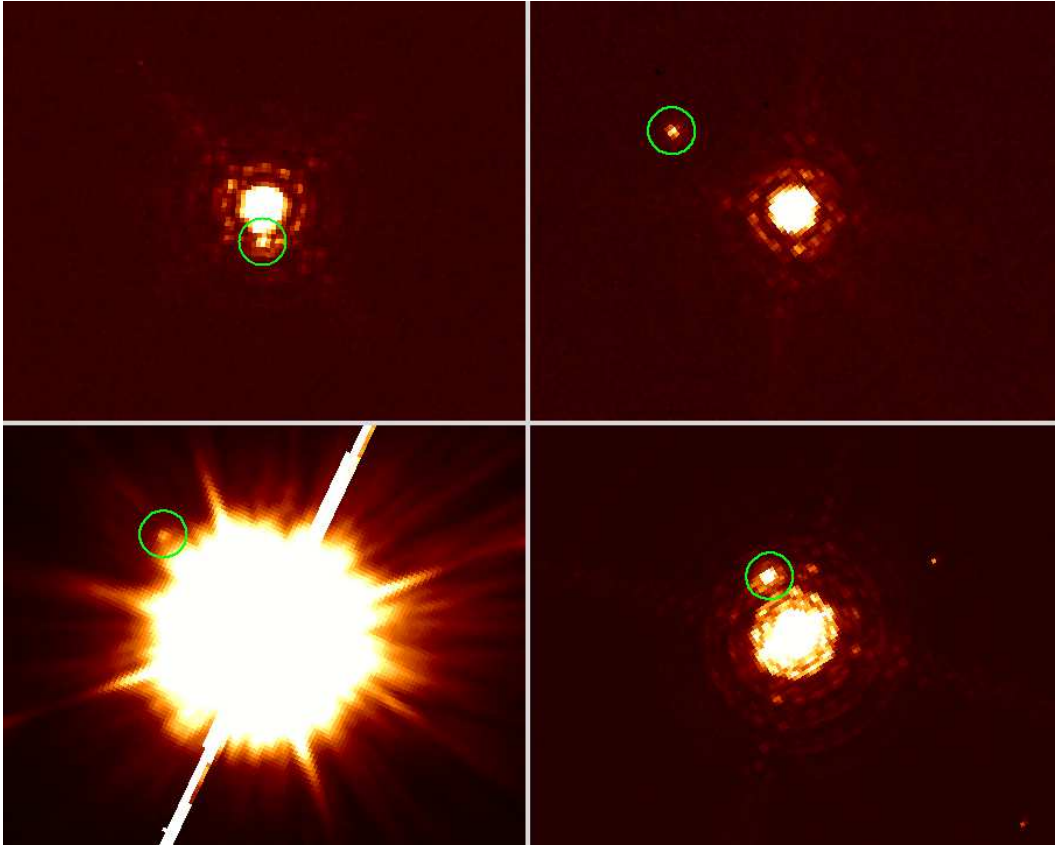}
\caption{
False-color renditions of \HST\/ images of \mucas. Each frame is $4\farcs0$ high
and has north at the top and east on the left. The companion star \mucas~B is
marked with a green circle in each image. {\it Top row}: WFPC2/PC images in the
F953N filter taken in 1999 ({\it left}) and 2007 ({\it right}); separations
$0\farcs349$ and $1\farcs389$. {\it Bottom row}: WFC3/UVIS images in F225W ({\it
left}) taken in 2012, and in F953N taken in 2016 ({\it right}); separations
$1\farcs333$ and $0\farcs607$.
\label{fig:mucasmosaic}
}
\end{figure*}

To give an impression of the images obtained with the three different camera setups, we show false-color renditions of typical frames in Figure~\ref{fig:mucasmosaic}. The companion \mucas~B is marked with green circles. The top two images were taken with WFPC2/PC and the F953N filter in 1999 and 2007, showing the companion lying on an Airy ring of the primary in the
first frame, and well separated in the second. The bottom left frame was taken in 2012 with WFC3/UVIS in the ultraviolet F225W filter, in which the dM companion is relatively very faint. At the bottom right is a WFC3/UVIS F953N frame obtained in 2016.

\section{{\em HST} Astrometric Analysis\label{sec:analysis}}

For the astrometric measurements of separation and position angle (PA) for the \mucas\ system, we have three sets of \HST\/ images. These are (1)~WFPC2/PC frames in the F953N filter (1997--2007); (2)~WFC3/UVIS frames in F225W (2010--2015); and (3)~WFC3/UVIS frames in F953N (2016). 

\subsection{WFPC2 Images in F953N\label{sec:wfpc2f953n}}

In the WFPC2 F953N bandpass, we obtained a total of 312 individual frames over
the 20 visits to \mucas. We used a nominal gain of 15 electrons per data number
and exposure times such that \mucas~A would just approach saturation. In only
seven cases was there actually a saturated pixel at the center of A's image, and
we discarded those frames from further analysis. In this long-wavelength
bandpass there were well-detected images of the B~companion (see Figure~\ref{fig:mucasmosaic} top
row). 

For the astrometric centroiding, we used a technique of point-spread-function
(PSF) fitting. Our procedure was nearly identical to that described in detail by
B15 for our analysis of unsaturated frames of Procyon, so we give only a brief
description here. The primary difference was that the \mucas\ frames were taken
in a different filter than we used for Procyon (F218W)\null. 

To determine a highly over-sampled PSF, we stacked the 305 frames using a
preliminary set of centroid estimates for the A component. By fitting to this
initial PSF, we updated the centroid positions of A\null. Then the refined
positions were used to create a new PSF\null. Iterating this procedure five
times led to excellent convergence. Using this final PSF, in the form of a
$5\times5$ array without the corners (i.e., 21 pixels), we determined each of
the A-image centroid positions.

To centroid the companion's images in the same frames, we first had to remove
light due to the wings and Airy rings of the primary star, which is particularly
important at small angular separations (e.g., the upper left frame in Figure~1).
This was done by defining a large-scale PSF based on the images of A, extending
out to the largest separation reached by B (suppressing pixels lying near B in
each individual frame before combining all of the images)\null. We then
subtracted this PSF from the frames, and then simply determined the positions of
B by fitting the same PSF employed for A\null. As described in B15, the
resulting $x,y$ positions were corrected for the WFPC2 34th-row anomaly, and for
geometric distortion.

In order to convert the adjusted $x,y$ positions to seconds of arc, we need the plate scale for WFPC2 F953N images. Unfortunately, this rarely used filter does not have a primary scale calibration. However, we found that the published plate scales for well-calibrated WFPC2 filters are strongly correlated with the index of refraction of MgF$_2$ at the effective wavelengths of the bandpasses. (This is due to the use of MgF$_2$ transmission optics in the camera.) Applying this relationship, we adopted a nominal plate scale of $0\farcs045575\,\rm pix^{-1}$.

The scale for each image was then very slightly adjusted for differential 
velocity aberration, using the image-header keyword {\tt VAFACTOR}\null. ``Breathing'' of
the telescope tube induces small changes in focus, and thus changes in the PSF along with minor
changes to the large-scale geometric distortion \citep[see][]{Gilliland2005}.  These
telescope responses tend to vary over the orbital visibility period.
Since our observations at each epoch used a full visibility period, these effects
will be somewhat averaged out.  Remaining uncorrected residuals resulting from these effects 
are a likely source for the small remaining scatter in our astrometry, as discussed below (\S\ref{sec:3rdbody}). 

Finally,
the orientation of each image on the sky was obtained from the {\tt ORIENTAT} keyword in the image headers, which has an uncertainty of about $\pm\!0\fdg03$ (see B15).
A small number of discordant measures were dropped (mostly due to cosmic-ray hits within the image of either star, or detector artifacts), and then the determinations at each epoch were combined into averages, with the uncertainties calculated from the standard errors of the means. 

An issue emerged when we began to make orbital solutions for the binary using the WFPC2 astrometry. Over the last several years of WFPC2 data, there were increasingly large residuals, with alternating signs, for observations spaced about six months apart. Since the spacecraft roll angles differed by about $180^\circ$ for successive visits, these offsets are plausibly attributable to the effect of charge-transfer inefficiency (CTI) in the  WFPC2 detectors. The amount of CTI increases with time, as the CCDs are exposed to the space environment. CTI leads to some of the charge in a stellar image falling behind as the image is read out, producing faint ``tails'' adjacent to the image, and thus slightly displacing its mean position in the detector $y$ direction. We derived an approximate empirical time-dependent correction for the CTI effect, as described in more detail in Appendix~\ref{sec:appendixa}, and applied it to all of the WFPC2 measurements. The final WFPC2 astrometric results are presented in the first 20 lines of Table~\ref{table:hstastrometry}.

\subsection{WFC3 Images in F225W\label{sec:wfc3f225w}}

In this series of observations, we obtained short and long exposures at each dithered telescope position. The image of A was unsaturated in the short exposures, and B was well detected in the long ones. We began by obtaining the drizzle-combined images (``drz'' frames) from the standard \HST\/ pipeline.\footnote{\url{https://archive.stsci.edu/hst}} Unlike the WFPC2 images, these frames have already been corrected for geometric distortion. They have cosmic rays removed, and the plate scale is given in the image headers. At each epoch, we have two pairs of short and long combined exposures.

We then proceeded similarly to the WFPC2 analysis described above. We determined an oversampled PSF by combining all of the \mucas\ frames, as well as a selection of 22 F225W observations of other bright stars available in the archive. These had all been taken in the same subarray and UVIS2 chip as our \mucas\ observations (except for our first WFC3 visit in UVIS1). Most of the archival frames are of white dwarfs that are much bluer than the components of \mucas, but we saw no evidence of a color term in the PSF\null. As for the WFPC2 frames, the PSF determination converged after a few iterations. We then used PSF fitting to determine the final positions for component~A.

In the long-exposure frames, the faint B companion is embedded in the bright wings of A (see Figure~1 lower-left panel). We determined a large-scale PSF from the observations of A, and subtracted it from the images before measuring the position of B, using the same oversampled PSF determined above. 

Our first WFC3 observations, taken in early 2010, present a special problem: they used the UVIS1 chip, for which there are insufficient observations useful for determining an oversampled PSF\null. We therefore used the UVIS2 PSF for the astrometry of these frames.


When we carried out our initial orbital fits to the \HST\/ data, the WFC3+F225W measurements stood out as having unusually large residuals (up to about 40~mas), compared to those from the earlier WFPC2 series, and the final WFC3 observation described below. Upon investigation, we eventually realized that this problem arises because of a small amount of chromatic aberration in the WFC3 camera, combined with the fact that the F225W filter has a significant red leak. The result is that most of the light detected from the very red \mucas~B is actually transmitted through the red leak, rather than the main bandpass of the UV filter, which transmits most of the light of the primary star. Thus the image of \mucas~B is slightly displaced relative to that of the bluer A component. We were able to derive an approximate correction for this effect, as discussed in detail in Appendix~\ref{sec:appendixb}\null. 

As with WFPC2, as discussed in the previous subsection and in Appendix~\ref{sec:appendixa}, WFC3 has shown a progressive increase of CTI with time,
potentially contributing errors to our position measurement of the very
faint \mucas~B relative to the much brighter A\null.  Our astrometric analyses
were performed using the ``drz'' image products provided by the STScI pipeline; the ``drc'' products that
additionally have been corrected at the pixel level for CTI were not
available at the time of our analyses.  We have subsequently compared the drc and drz images, and they do not show discernible shifts of the position of B in our data.  We also performed
an empirical search for CTI-induced position shifts, as we did for 
WFPC2, but did not find a significant correlation of the $x, y$ residuals 
relative to a preliminary orbit fit as a function of time.  In the WFC3 F225W images, the B
component is well within an extended halo of light from the much 
brighter~A, producing a local sky background of several hundred electrons per pixel at its position. This background likely suppresses any significant CTI losses. Thus we did not make any corrections for CTI in the drz images.

The next six lines in Table~\ref{table:hstastrometry} contain the results of these measurements, adjusted for differential chromatic aberration as described above. The uncertainties were calculated based on the internal scatter of the pairs of measurements at each epoch, combined in quadrature with an estimated error of $0\farcs0007$ from telescope pointing drift between the short and long exposures (see B15 for details), and the uncertainty in telescope orientation described above. We have not attempted to include the additional systematic uncertainties due to the approximate nature of the aberration correction. Because of this, we will give the F225W measurements a lower weight than the other determinations when we calculate an orbital fit below.

\subsection{WFC3 Images in F953N}

Our final 2016 observations of \mucas\ were made with the WFC3's long-wavelength
F953N filter. We obtained dithered images with short (0.5~s), medium (2.5~s),
and long (200~s) exposures. The short exposures proved to be a mixture of
saturated and unsaturated images of A, and were discarded. In the
medium-exposure frames, A is saturated, and in the long-exposure images both
stars are saturated.  With the F953N exposures, we have the advantage that the
same filter was used for our studies of Procyon (B15) and Sirius (B17). Thus we can use very similar reduction techniques. As those papers describe, we employed two different methods for centroiding the stellar images. One used PSF fitting, based on selected regions in the PSF in the unsaturated outskirts. The other used the diffraction spikes in the overexposed images, taking their inferred intersection point as the centroid. As noted in B15 and B17, the two methods give results that agree well. In the final line in Table~\ref{table:hstastrometry} we give the average
separation and PA obtained from the two methods. 

As with the WFC3 observations in F225W, the well-exposed image of B
in the F953N frames sits on top of hundreds of electrons from the
nearby A\null.  Fitting the location of B in the drc images shows no
difference from those we derived using drz frames.  The high sky 
background, coupled with well-exposed B images for the F953N 
exposures, likely suppressed any significant CTI losses.

\begin{deluxetable}{lccc}
\tablewidth{0 pt}
\tabletypesize{\footnotesize}
\tablecaption{{\em HST\/} Astrometric Measurements of \mucas~B Relative to \mucas~A
\label{table:hstastrometry}
}
\tablehead{
\colhead{UT Date} &
\colhead{Besselian} &
\colhead{Separation} &
\colhead{J2000 Position} \\
\colhead{} &
\colhead{Date} &
\colhead{[arcsec]} &
\colhead{Angle\tablenotemark{a} [$^\circ$]} 
}
\startdata
\noalign{\smallskip}
\multispan4{\hfil WFPC2/PC Frames, F953N Filter\tablenotemark{b} \hfil} \\
1997 Jul 04 & 1997.5057 & $0.4191\pm0.0010$ & $226.493\pm0.092$ \\  
1998 Jan 02 & 1998.0047 & $0.4454\pm0.0009$ & $214.181\pm0.066$ \\ 
1998 Jul 22 & 1998.5544 & $0.4077\pm0.0002$ & $200.497\pm0.075$ \\ 
1999 Feb 28 & 1999.1612 & $0.3490\pm0.0004$ & $179.581\pm0.087$ \\ 
1999 Aug 04 & 1999.5906 & $0.3151\pm0.0007$ & $160.771\pm0.101$ \\ 
2000 Feb 01 & 2000.0868 & $0.3221\pm0.0006$ & $137.050\pm0.071$ \\ 
2000 Jul 15 & 2000.5390 & $0.3632\pm0.0005$ & $118.183\pm0.064$ \\ 
2001 Jan 15 & 2001.0406 & $0.4313\pm0.0004$ & $102.617\pm0.067$ \\ 
2001 Jul 30 & 2001.5773 & $0.5237\pm0.0003$ & $ 91.325\pm0.057$ \\ 
2002 Jan 17 & 2002.0476 & $0.6103\pm0.0004$ & $ 84.393\pm0.040$ \\ 
2002 Aug 05 & 2002.5934 & $0.7102\pm0.0003$ & $ 78.543\pm0.059$ \\ 
2003 Feb 11 & 2003.1144 & $0.8028\pm0.0003$ & $ 74.109\pm0.036$ \\ 
2003 Aug 05 & 2003.5942 & $0.8860\pm0.0006$ & $ 70.993\pm0.036$ \\ 
2004 Jan 29 & 2004.0772 & $0.9637\pm0.0002$ & $ 68.262\pm0.037$ \\ 
2004 Aug 08 & 2004.6039 & $1.0429\pm0.0006$ & $ 65.927\pm0.040$ \\ 
2005 Jan 15 & 2005.0412 & $1.1071\pm0.0003$ & $ 64.020\pm0.035$ \\ 
2005 Aug 13 & 2005.6175 & $1.1826\pm0.0005$ & $ 61.937\pm0.035$ \\ 
2006 Jan 30 & 2006.0815 & $1.2368\pm0.0003$ & $ 60.405\pm0.035$ \\ 
2006 Sep 26 & 2006.7400 & $1.3054\pm0.0004$ & $ 58.568\pm0.034$ \\ 
2007 Oct 17 & 2007.7933 & $1.3931\pm0.0003$ & $ 55.798\pm0.032$ \\ 
\noalign{\smallskip}
\multispan4{\hfil WFC3/UVIS Frames, F225W Filter\tablenotemark{c} \hfil} \\
2010 Jan 09 & 2010.0236 & $1.4795\pm0.0037$ & $ 50.711\pm0.072$ \\ 
2010 Dec 03 & 2010.9222 & $1.4625\pm0.0058$ & $ 48.697\pm0.284$ \\ 
2011 Dec 05 & 2011.9264 & $1.4166\pm0.0018$ & $ 46.204\pm0.076$ \\ 
2012 Dec 02 & 2012.9204 & $1.3326\pm0.0011$ & $ 43.439\pm0.099$ \\ 
2013 Oct 25 & 2013.8179 & $1.2365\pm0.0008$ & $ 40.960\pm0.123$ \\ 
2015 Jan 06 & 2015.0150 & $0.9969\pm0.0007$ & $ 37.122\pm0.082$ \\ 
\noalign{\smallskip}
\multispan4{\hfil WFC3/UVIS Frames, F953N Filter \hfil} \\
2016 Jul 11 & 2016.5261 & $0.6069\pm0.0028$ & $ 25.450\pm0.280$ \\ 
\enddata
\tablenotetext{a}{Note that the PAs are referred to the equator of J2000, not to the equator of observation epoch as is the usual practice for ground-based visual-binary measurements.}
\tablenotetext{b}{Corrected for charge-transfer inefficiency, as described in \S\ref{sec:wfpc2f953n} and Appendix~\ref{sec:appendixa}.}
\tablenotetext{c}{Corrected for chromatic aberration, as described in \S\ref{sec:wfc3f225w} and Appendix~\ref{sec:appendixb}.}
\end{deluxetable}


\section{Ground-based Measurements and Parallax}

\subsection{Astrometry of \mucas~B}

Although the available ground-based astrometric measurements of \mucas~B relative to A generally do not have the precision of the \HST\/ data, they cover more than twice the time interval. Thus they are useful for constraining orbital parameters, especially the orbital period.
Table~\ref{table:groundastrometry} lists the published ground-based astrometric observations of \mucas~B of which we are aware, along with one unpublished measurement from a private communication.

\begin{deluxetable}{lccc}
\tablewidth{0 pt}
\tabletypesize{\footnotesize}
\tablecaption{Ground-based Astrometric Measurements of \mucas~B  Relative to \mucas~A
\label{table:groundastrometry}
}
\tablehead{
\colhead{Besselian} &
\colhead{Separation} &
\colhead{Position} &
\colhead{Reference\tablenotemark{a}}  \\
\colhead{Date} &
\colhead{[arcsec]} &
\colhead{Angle  [$^\circ$]} &
\colhead{}  
}
\startdata
1973.787   & $ 0.35   \pm 0.04  $ & $ 24   \pm 3   $ & (1)  \\
1974.650   & $ 0.23   \pm 0.01  $ & $333.9 \pm 2   $ & (2)  \\
1983.20    & $ 0.98   \pm 0.024 $ & $ 55.8 \pm 0.7 $ & (3)  \\
1983.494   & $ 0.93   \pm 0.06  $ & $ 224.3 \pm 2.6 $ & (4)  \\
1983.7072  & $ 1.074  \pm 0.042 $ & $ 61.9 \pm 3.0 $ & (5)  \\
1984.126   & $ 1.118  \pm 0.023 $ & $ 63   \pm 2   $ & (6)  \\
1984.9132  & $ 1.251  \pm 0.030 $ & $ 59.3 \pm 1.3 $ & (5)  \\
1985.0842  & $ 1.320  \pm 0.027 $ & $ 60.6 \pm 1.1 $ & (5)  \\
1985.8448  & $ 1.425  \pm 0.016 $ & $ 59.47\pm 0.51$ & (5)  \\
1990.6836  & $ 1.36   \pm 0.076 $ & $ 48.2 \pm 3.1 $ & (7)  \\
1991.7268  & $ 1.41   \pm 0.051 $ & $ 48.6 \pm 6.8 $ & (7)  \\
1994.6563  & $ 0.73   \pm 0.02  $ & $ 28.5 \pm 0.7 $ & (8)  \\
1994.8069  & $ 0.66   \pm 0.02  $ & $ 27.0 \pm 0.8 $ & (8)  \\
2003.5663  & $ 0.86		$ & $ 70.2	   $             & (9)  \\
2004.6632  & $ 1.042  \pm 0.020 $ & $ 65.0 \pm 1.3 $ & (10)  \\
2014.7581  & $ 1.0707 \pm 0.0036$ & $ 38.3 \pm 0.51$ & (11) \\ 
2014.7581  & $ 1.0727 \pm 0.0036$ & $ 38.1 \pm 0.51$ & (11) \\ 	
2015.5448  & $ 0.9018 \pm 0.0043$ & $ 34.5 \pm 1.0 $ & (12) \\
2015.5448  & $ 0.8991 \pm 0.0043$ & $ 32.0 \pm 1.0 $ & (12) \\
2016.0337  & $ 0.7560 \pm 0.0037$ & $ 30.9 \pm 1.0 $ & (12) \\
2016.0337  & $ 0.7510 \pm 0.0037$ & $ 31.0 \pm 1.0 $ & (12) \\
2016.0474  & $ 0.7513 \pm 0.0037$ & $ 30.6 \pm 1.0 $ & (12) \\
2016.0474  & $ 0.7589 \pm 0.0037$ & $ 27.7 \pm 1.0 $ & (12) \\
\enddata					    
\tablenotetext{a}{References: (1) Wickes \& Dicke 1974; (2) Wickes 1975; (3) McCarthy 1984; (4) \citet{Pierce1985}; (5) Haywood et al.\ 1992; (6) Karovska et al.\ 1986; (7) McCarthy et al.\ 1993; (8) Drummond et al.\ 1995; (9) L.~Roberts, Palomar AO system, private communication; (10) Christou \& Drummond 2006; (11 and 12) \citet{Horch2015, Horch2019}; observations at each epoch were made in two different bandpasses.}  
\end{deluxetable}

\subsection{Parallax\label{sec:parallax}}

\mucas\ is not included in the recent \Gaia\/ Data Release~2 (DR2) \citep{Gaia2018}, likely because of the star's brightness and large proper motion. However, parallax measurements are available from several earlier studies. \citet{Lippincott1964} list their own measurement, along with three earlier determinations, but since all of their stated uncertainties are relatively large compared to more recent values we did not utilize them in our study. 

The five parallax determinations that we considered are listed in the first five lines in Table~\ref{table:parallax}: (1)~\citet{Lippincott1981} obtained the parallax from measurements of photographic plates taken at the Sproul Observatory on 215 nights between 1937 and 1980, converted from relative to absolute using a statistical mean parallax for the reference stars. Since precise parallaxes are now available for each of the background stars from \Gaia\/ DR2, we calculated the mean of these and made a (small) adjustment to her result. She had assumed a mean parallax of $0\farcs0041$ but the DR2 mean is $0\farcs0019$. (2)~\citet{Russell1984} measured the parallax using 371 plates from the Allegheny Observatory taken between 1933 and 1978. We again made a small adjustment to their result, based on the mean \Gaia\/ parallaxes for their reference stars of $0\farcs0025$ vs.\ their assumed $0\farcs0030$. (3)~\citet{Harrington1993} measured 68 plates obtained at the U.S. Naval Observatory between 1984 and 1990, and similarly adjusted from relative to absolute using a statistical algorithm. Here the adjustment based on the mean \Gaia\/ reference-star parallaxes is very small, $0\farcs0017$ as compared to their $0\farcs0015$. (4)~\citet{Heintz1994} presented the final Sproul photographic results, from 251 observations over 55~years. He did not give details of the reference stars, so we assumed they were the same as used by Lippincott and we applied the same DR2-based correction to absolute. (5)~The absolute parallax was measured by the \Hipp\/ mission \citep{vanLeeuwen2007}. 

The five results are in good agreement. Omitting the earlier Sproul measurement as being superseded by the later one, we adopt the weighted mean of the remaining four measurements (which is very close to the \Hipp\/ value), as given in the final line of Table~\ref{table:parallax}.


\begin{deluxetable}{lcl}
\tablewidth{0 pt}
\tablecaption{Parallax of \mucas
\label{table:parallax}}
\tablehead{
\colhead{Source} & 
\colhead{Parallax [arcsec]}  &
\colhead{Reference} 
}
\startdata
Sproul      & $0.1318 \pm 0.0011$\tablenotemark{a} & \citet{Lippincott1981} \\
Allegheny   & $0.1363 \pm 0.0033$\tablenotemark{a} & \citet{Russell1984} \\
USNO        & $0.1326 \pm 0.0023$\tablenotemark{a} & \citet{Harrington1993} \\
Sproul      & $0.1329 \pm 0.0017$\tablenotemark{a} & \citet{Heintz1994} \\
\Hipp\/     & $0.13238 \pm 0.00082 $ & \citet{vanLeeuwen2007}     \\
\noalign{\smallskip}
Weighted mean & $ 0.13266 \pm 0.00069  $ & Adopted\tablenotemark{b}       \\
\noalign{\smallskip}
\enddata
\tablenotetext{a}{Adjusted for mean \Gaia\/ DR2 parallax of reference stars; see text.}
\tablenotetext{b}{\citet{Lippincott1981} value not included in the mean; see text.}
\end{deluxetable}

\subsection{Photocenter Motion of \mucas~A}

To obtain the individual masses of the two components we require the semimajor axis of the absolute orbital motion of \mucas~A\null. Because of the large magnitude difference between the components, we take the photocenter of the system to represent this motion. Measurements of the photocenter motion were made in the parallax studies of \citet{Lippincott1981} and \citet{Russell1984}. Lippincott listed normal points for her measurements of the photocentric orbit (her Table~2). \citet{Russell1984} did not tabulate their individual measurements, but J.~Russell had provided them privately to \citet{Drummond1995}. Through the kindness of J.~Russell and J.~Drummond, these measurements were communicated to us. Because they have not been published previously, we list them in Table~\ref{table:russell}. These data will be included as input to the final orbital solution described below.

\begin{deluxetable}{lcc}
\tablewidth{6in}
\tablecaption{Photocenter Motion for \mucas~A Measured by \citet{Russell1984}\tablenotemark{a}
\label{table:russell}}
\tablehead{
\colhead{Epoch} & 
\colhead{\hspace{.1in}Offset}\hspace{.2in}  &
\colhead{Position}\hspace{.2in} \\ 
\colhead{} & 
\colhead{\hspace{.1in}[arcsec]}\hspace{.2in}  &
\colhead{Angle [$^\circ$]}\hspace{.2in}  
}
\startdata
1933.0160 & 0.1471 &  38.1238 \\
1933.7960 & 0.0655 &  10.6570 \\
1934.6460 & 0.0664 & 309.5121 \\
1935.6290 & 0.0751 & 284.2946 \\
1937.6650 & 0.1494 & 257.3742 \\
1938.9510 & 0.1500 & 272.5285 \\
1939.6970 & 0.2092 & 243.6260 \\
1940.6360 & 0.2395 & 248.4918 \\
1942.9270 & 0.2827 & 231.6788 \\
1943.8620 & 0.2651 & 235.2058 \\
1944.6930 & 0.2654 & 231.6809 \\
1945.7420 & 0.2694 & 230.8118 \\
1946.7660 & 0.2662 & 232.2710 \\
1947.8000 & 0.2540 & 226.8881 \\
1948.7960 & 0.2193 & 225.9624 \\
1949.8390 & 0.1795 & 224.9168 \\
1950.8150 & 0.1475 & 213.4607 \\
1951.7580 & 0.1158 & 213.0443 \\
1952.7880 & 0.0615 & 142.6767 \\
1953.7200 & 0.0684 &  50.9950 \\
1954.7770 & 0.0903 &  25.8460 \\
1955.6750 & 0.0739 &   8.4718 \\
1957.8580 & 0.0832 & 293.9446 \\
1964.8380 & 0.2305 & 236.1935 \\
1965.7650 & 0.2628 & 240.7150 \\
1966.9150 & 0.2556 & 239.1292 \\
1968.6880 & 0.2713 & 229.5558 \\
1969.7280 & 0.2589 & 228.6600 \\
1971.9450 & 0.1330 & 218.8055 \\
1972.6250 & 0.1959 & 226.5597 \\
1975.8830 & 0.0829 &  33.0381 \\
1976.7680 & 0.0772 &  43.5241 \\
1977.6930 & 0.0437 &   5.3173 \\
1978.8030 & 0.0366 & 359.9587 \\
\enddata
\tablenotetext{a}{Previously unpublished data, kindly communicated by J.~Drummond and J.~Russell.}
\end{deluxetable}

\subsection{Radial Velocities\label{sec:radvels}}

RV measurements potentially provide useful constraints on the orbital solution, especially since they cover more than a century. They also resolve the ambiguity as to the orientation of the orbit (i.e., which star is in front). We compiled the RV data published in the following papers:
(1)~\citet{Worek1977}: 100 photographic measurements, 1900--1976.
(2)~\citet{Abt1980}: 3 photographic measurements, 1967--1975.
(3)~\citet{Beavers1986}: 22 RV spectrometer measurements, 1976--1983.
(4)~\citet{Abt1987}: 12 CCD measurements, 1984--1985.
(5)~\citet{Abt2006}: 24 CCD measurements, 2000--2003.
(6)~\citet{Agati2015}: 45 CORAVEL measurements, 1977--1999, including a re-reduction of data published earlier by \citet{Jasnie1988} and \citet{Duquennoy1991}.

\section{Elements of the Relative Visual Orbit of \mucas~B}

\subsection{Orbital Solution\label{sec:solution}}

Our determination of the orbital elements largely follows the procedures described in detail for Procyon and Sirius by B15 and B17. We describe the main points of the fitting method below.

The first step was to adjust all of the measurements, \HST\/ and ground-based, to the J2000 standard equator and epoch. We used the formulations given by \citet{vandenBos1964} in order to correct for (1)~precession (except for the \HST\/ measures, which are already in the J2000 frame), (2)~the change in direction to north due to proper motion, (3)~the changing viewing angle of the three-dimensional orbit due to proper motion, and (4)~the steadily decreasing distance of the system due to RV\null. All of these corrections are small relative to the observational uncertainties for the ground-based data, and are also small for the \HST\/ data because their epochs are all so close to 2000.0.


We determined elements for the relative visual orbit and photocenter motion via an eight-parameter fit to the combined set of J2000-corrected \HST\/ and ground-based measurements of the B-A separation and PA (Tables~\ref{table:hstastrometry} and \ref{table:groundastrometry}, with adjustments applied), and of the photocenter motion of~A \citep[][and our Table~\ref{table:russell}]{Lippincott1981}.  This fit employed a Newton-Raphson method to minimize the $\chi^2$ between the measured and fitted positions, by calculating a first-order Taylor expansion for the equations of orbital motion. The procedure results in a solution for the period $P$, time of periastron passage $T_0$, eccentricity $e$, semimajor axis $a$, inclination $i$, PA of the line of nodes $\Omega$, argument of periastron $\omega_B$ as referenced to $\mu$~Cas~B, and the semimajor axis of the photocenter motion $a_A$.  

Before computing the joint fit to all data, we fit an orbit to each set of measurements independently, and scaled the uncertainties in order to force the reduced $\chi^2_\nu$ to unity.  We scaled the error estimates for WFPC2 by a factor of 2.7, WFC3 by 7.3, and the ground-based measurements by 1.8, compared with the values listed in Tables~\ref{table:hstastrometry} and~\ref{table:groundastrometry}. For the ground-based observations we deleted the 1983.20 measurement, because it was $\sim$4$\sigma$ discrepant from the initial orbit fit.
The large uncertainty scale factor found for the WFC3 data is perhaps not surprising, given the approximate nature of the chromatic-aberration corrections described in Appendix~\ref{sec:appendixb}\null. The smaller scaling for WFPC2 probably reflects remaining systematic errors due to telescope breathing and CTI, as discussed in \S\ref{sec:wfpc2f953n}. For the photocenter motion, we assumed equal uncertainties in separation of $0\farcs0175$ for all of the measurements, and scaled the uncertainties in PA to produce equal uncertainties in right ascension and declination.  The final orbital parameters determined from the joint fit to the visual orbit and photocenter motion are given in Table~\ref{table:elements}.

\begin{figure*}[ht]
\centering
\includegraphics[width=3.9in]{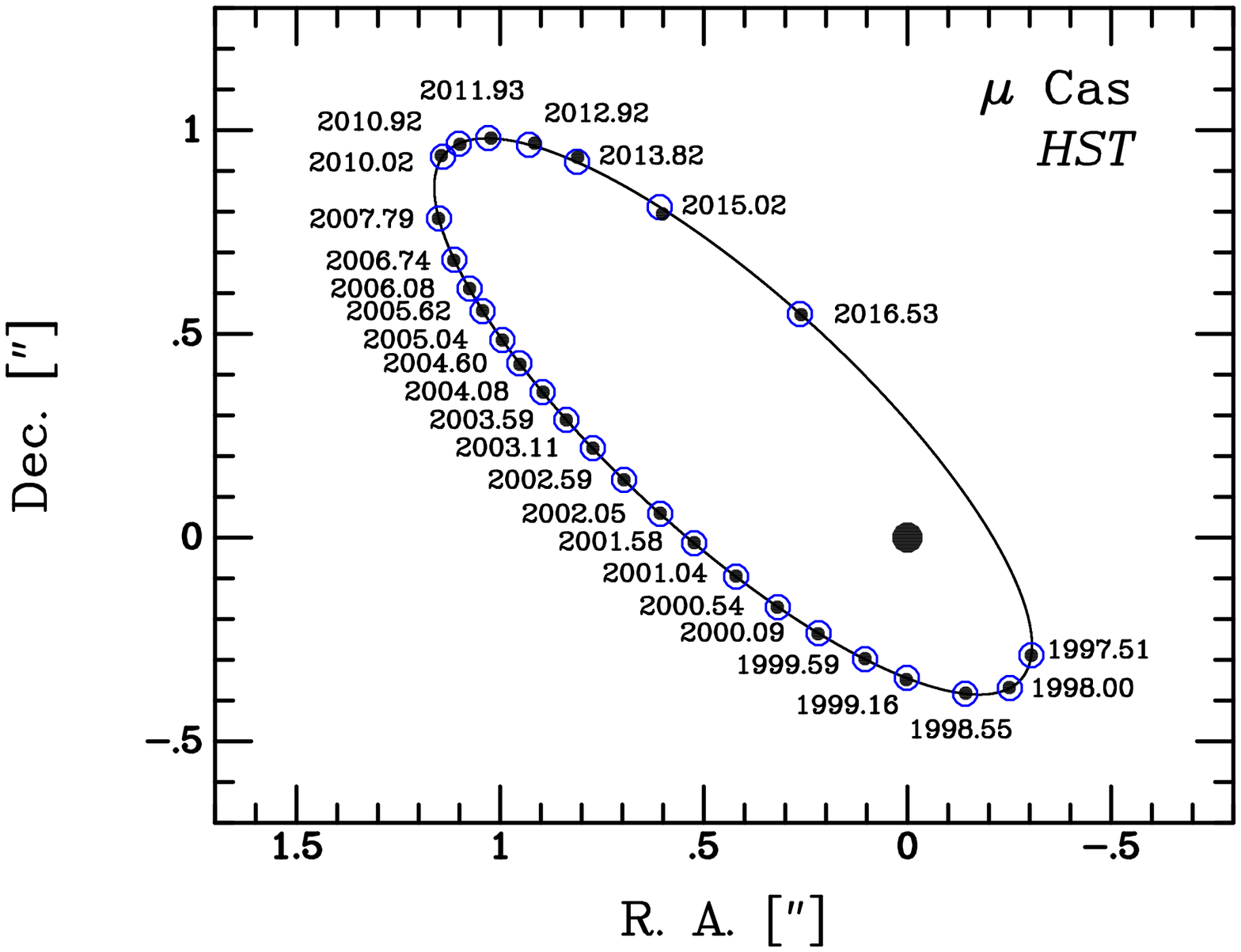}
\vskip0.25in
\includegraphics[width=3.9in]{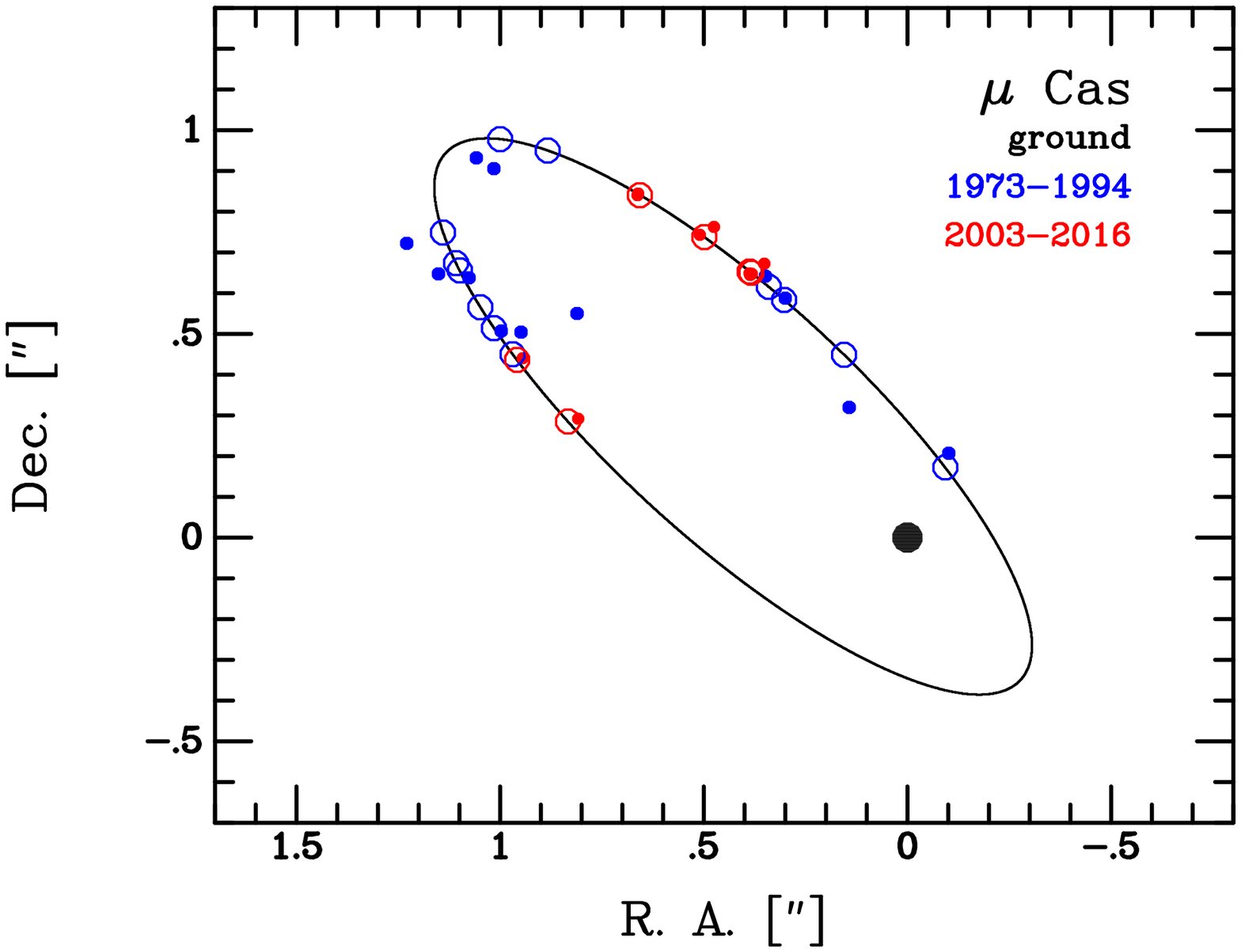}
\caption{
{\it Top panel:} \HST\/ observations of the orbit of \mucas~B relative to A (which is plotted as a large black point at the origin). Filled black circles show the \HST\/ measurements (listed in Table~\ref{table:hstastrometry}), each one labelled with the date of observation. The black ellipse plots our orbital fit from \S\ref{sec:solution}. Open blue circles mark the predicted positions from the orbital fit.
{\it Bottom panel:} Ground-based measurements (listed in Table~\ref{table:groundastrometry}). The black ellipse is our orbital fit from the top panel. Blue filled circles plot the observations from 1973 to 1994, and red filled circles show the measurements from 2003 to 2016. The open blue and red circles mark the corresponding predicted positions based on our orbital fit. 
\label{fig:orbits}
}
\end{figure*}

Figure~\ref{fig:orbits} depicts the orbit of \mucas~B\null. The top panel plots the positions of B relative to A as measured by \HST\null, and the bottom panel shows the ground-based measurements. In both panels the black ellipse shows our orbital fit. In the top panel the filled black circles mark the \HST\/ measurements from Table~\ref{table:hstastrometry}, with the small adjustments described above applied. The open blue circles are the predicted positions from our orbital parameters. The fit agrees well with the WFPC2 data (1997--2007) at the scale of the figure, as does the final WFC3 observation in 2016. For the WFC3 F225W observations, 2010--2015, there are evident small departures from the fit, likely arising from uncertainties in the correction for chromatic aberration, as just noted above.

The bottom panel of Figure~\ref{fig:orbits} plots the ground-based measurements from Table~\ref{table:groundastrometry}, again with the small adjustments applied. The blue filled circles mark the observations from 1973 through 1994, and the red filled circles those from 2003 to 2016. The open circles, with the same color-coding, show the corresponding ephemeris positions from our orbit solution. As can be seen, the early observations had significant errors. The 21st-century observations have noticeably smaller errors. 

\begin{deluxetable}{ll}
\tablewidth{0 pt}
\tablecaption{Elements of \mucas\ Visual Orbit (J2000)
\label{table:elements}}
\tablehead{
\colhead{Element} &
\colhead{Value} 
}
\startdata
Orbital period, $P$ [yr]                     &   21.568  $\pm$ 0.015  \\
Semimajor axis, $a$ [arcsec]                 &    0.9985 $\pm$ 0.0013 \\
Inclination, $i$ [$^\circ$]                  &  110.671   $\pm$ 0.064   \\
Position angle of node, $\Omega$ [$^\circ$]  &  223.868   $\pm$ 0.064   \\
Date of periastron passage, $T_0$ [yr]       & 1997.2235  $\pm$ 0.0067  \\
Eccentricity, $e$                            &    0.5885 $\pm$ 0.0011 \\
Longitude of periastron, $\omega_B$ [$^\circ$] &  330.37   $\pm$ 0.18   \\
Photocenter semimajor axis, $a_A$ [arcsec] & 0.1882 $\pm$ 0.0023 \\
\enddata
\end{deluxetable}

\subsection{Photocenter Orbit}

\begin{figure*}[ht]
\centering
\includegraphics[width=3.9in]{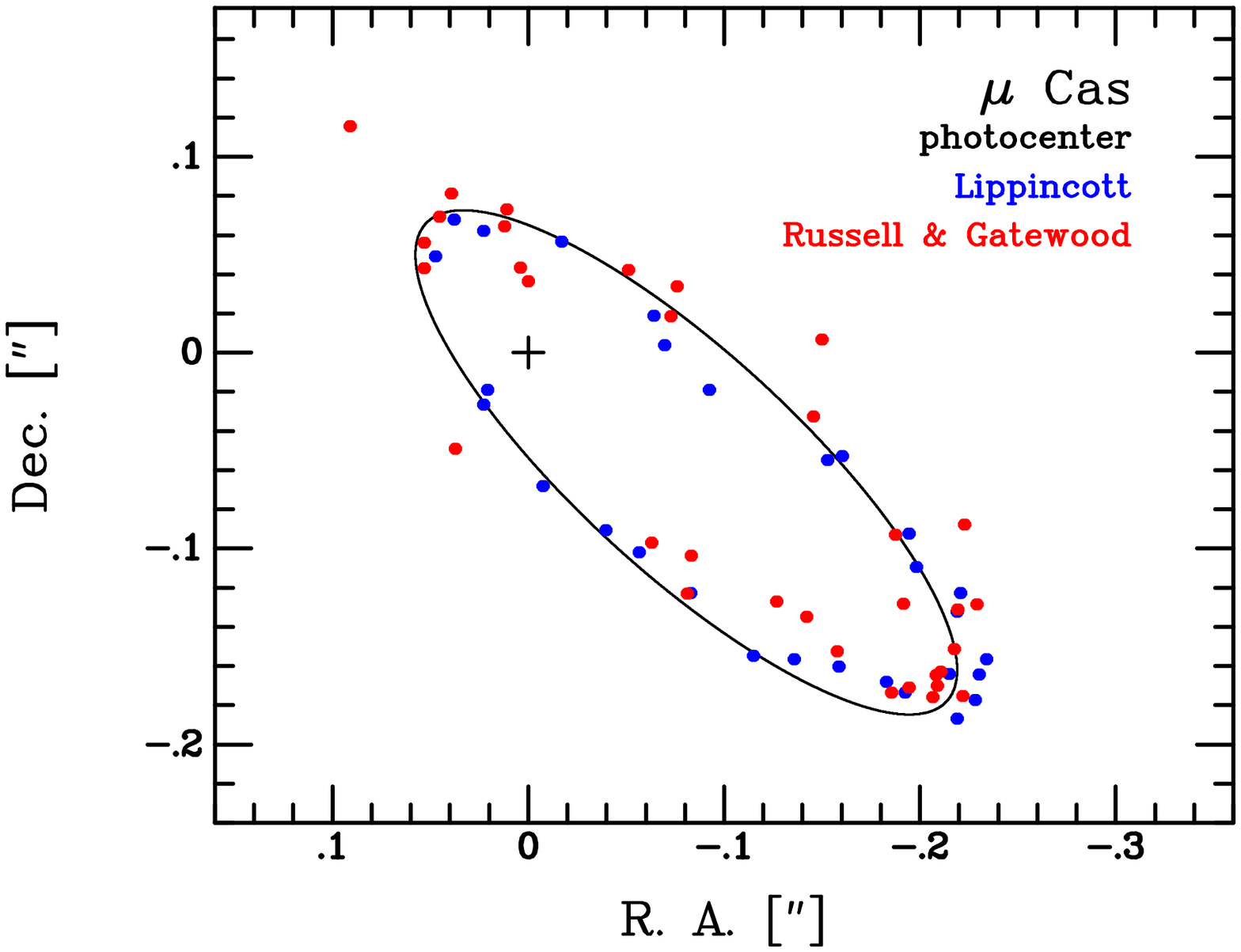}
\caption{
Offsets of the \mucas\ photocenter relative to the center of mass (plus sign), from the ground-based measurements of \citet{Lippincott1981} (filled blue circles) and \citet{Russell1984} (from our Table~\ref{table:russell}; filled red circles). The black ellipse shows the fit from our joint orbital solution.
\label{fig:photocenter}
}
\end{figure*}

Figure~\ref{fig:photocenter} plots the positions of the photocenter of \mucas\ relative to the center of mass. The filled blue and red circles are the Sproul and Allegheny measurements of \citet{Lippincott1981} and of \citet{Russell1984} (from our Table~\ref{table:russell}), respectively. The black ellipse is our orbital fit from our $\chi^2$ solution, with a semimajor axis of $a_A=0\farcs1882\pm0\farcs0023$. Lippincott obtained a value of $0\farcs1862\pm0\farcs0013$ from her data, and Russell \& Gatewood found $0\farcs1900\pm0\farcs0038$ from theirs. \citet{Drummond1995} carried out a joint solution from their own astrometry and combining both the Lippincott and Russell \& Gatewood photocenter motions, obtaining $a_A=0\farcs1908\pm0\farcs0043$. All of these earlier results are in reasonable agreement with our final value.

\goodbreak

\subsection{Radial-Velocity Curve}

We attempted to add the \mucas~A RV semiamplitude, $K_A$, and the center-of-mass RV, $\gamma$, as ninth and tenth parameters in a joint orbital fit. We first tried to use all of the RV measurements from the references quoted in \S\ref{sec:radvels}. However, it was apparent that the earlier, mostly photographic, RV data have significantly larger errors than the later values obtained with digital detectors, and that there are systematic offsets between different observatories. We then considered only the modern RV measurements of \citet{Abt2006} and \citet{Agati2015} (as was done by Agati et al.\ in their discussion of \mucas). This ten-parameter fit produced larger parameter uncertainties than the purely astrometric eight-parameter solution described above (Table~\ref{table:elements}). Moreover, this solution resulted in an RV semiamplitude of $K_A=2.43\pm 0.09\,\kms$. A value this large, when combined with the measured photocenter semimajor axis, $a_A$, implies a distance to the system about 22\% larger than given by the directly measured parallax. For these reasons, we will retain the orbital elements from the purely astrometric solution (Table~\ref{table:elements}) in the discussion below. 

\begin{figure*}[ht]
\centering
\includegraphics[width=4.5in]{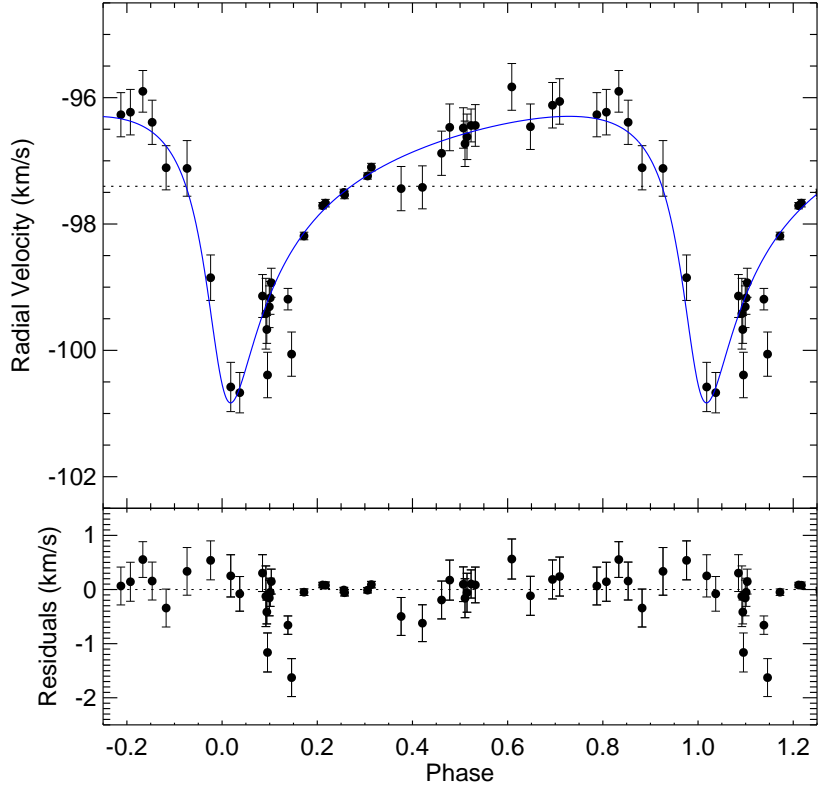} 
\caption{
{\it Top panel:} Radial velocities for \mucas~A (filled black circles) versus orbital phase. Measurements are from \citet{Abt2006} and \citet{Agati2015}. Velocities obtained within 7~days of each other have been combined into normal points. The blue line shows the velocity curve predicted based on the adopted parallax and our eight-parameter fit to the astrometry. {\it Bottom panel:} Residuals from the predicted velocities.
\label{fig:rvcurve}
}
\end{figure*}

The RV measurements nevertheless provide a useful check on our orbital solution. In the top panel of Figure~\ref{fig:rvcurve} we plot the \citet{Abt2006} and \citet{Agati2015} RV measurements versus orbital phase.\footnote{\citet{Abt2006} did not list uncertainties for their individual measurements; based on the discussion in their text, we adopted $\pm\!0.10\,\kms$ for each velocity.} In this plot, measurements obtained within 7~days of each other have been combined into normal points using weighted means. Based on the astrometric parameters and parallax, we predict a RV semiamplitude of $K_A = 2.27\pm 0.03\,\kms$. The blue line in the top panel shows the RV curve predicted by our orbital elements, where we have solved only for the center-of-mass RV, obtaining $\gamma=-97.40\pm0.03\,\kms$. The bottom panel plots the residuals of the observations versus the predicted values. The predictions appear to agree well with the measurements, especially the values with small uncertainties from \citet{Abt2006}. The larger $K_A$ that we found in the ten-parameter fit arose primarily from a few slightly discordant CORAVEL values around orbital phases 0.09--0.14; this $K_A$ differs by only $\sim$1.8$\sigma$ from the astrometrically predicted value.

\section{Dynamical Masses}

\subsection{Masses of \mucas~A and B\label{sec:dynmass}}

To calculate the dynamical masses of \mucas~A and~B, we employed the usual formula for the total system mass, $M = M_A+M_B = a^3/(\pi^3 \, P^2)$. The individual masses are then obtained using $M_A = M\,(1-a_A/a)$ and $M_B = M\,a_A/a\,$. In these equations the masses are in $M_\odot$, $a$ and $\pi$ are the semimajor axis and parallax in arcseconds, and $P$ is in years.

Table~\ref{table:masses} presents the dynamical masses given by \citet{Drummond1995}, \citet{Lebreton1999}, and \citet{Horch2019} in columns~2, 3, and~4, respectively. (The \citealt{Lebreton1999} value was simply the \citealt{Drummond1995} result adjusted to the {\it Hipparcos\/} parallax.) Our results are in the final column. They are in good agreement with the previous determinations, but the uncertainties are significantly smaller.

\begin{deluxetable*}{lcccc}
\tablewidth{0 pt}
\tabletypesize{\footnotesize}
\tablecaption{Dynamical Masses for the \mucas\ System 
\label{table:masses}}
\tablehead{
\colhead{Quantity} & 
\colhead{\citet{Drummond1995}} &
\colhead{\citet{Lebreton1999}} &
\colhead{\citet{Horch2019}} &
\colhead{This paper} 
}
\startdata
Total mass, $M_A+M_B$ & 
  $0.915 \pm 0.060 \, M_\odot$ & $\dots$ & $0.906 \pm 0.023\,M_\odot$ & $0.9168 \pm 0.0148 \, M_\odot$ \\
Mass of \mucas~A, $M_A$                 & 
  $0.742 \pm 0.059 \, M_\odot$ & $0.757 \pm 0.060\,M_\odot$ & $\dots$ & $0.7440 \pm 0.0122 \, M_\odot$ \\
Mass of \mucas~B, $M_B$                 & 
  $0.173 \pm 0.011 \, M_\odot$ & $\dots$ & $\dots$ & $0.1728 \pm 0.0035 \, M_\odot$ \\
\enddata
\end{deluxetable*}

\subsection{Error Budget}

Table~\ref{table:errorbudget} shows the contributions of the uncertainties of each orbital parameter to the overall uncertainties of the dynamical masses of \mucas~A and B\null. The uncertainty in the mass of A is almost entirely due to the error in the adopted parallax. For the mass of B, the uncertainty is due about equally to the uncertainties in the parallax and in the semimajor axis of the photocenter orbit, $a_A$. A more precise parallax from \Gaia\/ DR3 would provide a significant reduction in the uncertainties of the dynamical masses.

\begin{deluxetable*}{lllcc}
\tablewidth{0 pt}
\tablecaption{Error Budgets for \mucas\ System Dynamical Masses 
\label{table:errorbudget}
}
\tablehead{
\colhead{Quantity} &
\colhead{Value} &
\colhead{Uncertainty} &
\colhead{$\sigma(M_A)$ [$M_\odot$] } &
\colhead{$\sigma(M_B)$ [$M_\odot$] }
}
\startdata
Absolute parallax, $\pi$    & 0.13266  & $\pm$0.00069 arcsec & 0.0116 & 0.0027 \\
Semimajor axis, $a$         & 0.9985  & $\pm$0.0013 arcsec   & 0.0031 & 0.0004 \\
Semimajor axis for A, $a_A$ & 0.1882  & $\pm$0.0023 arcsec   & 0.0021 & 0.0021 \\
Period, $P$                 & 21.568 & $\pm$0.015 yr         & 0.0010 & 0.0002 \\
\noalign{\vskip0.1in}
Combined mass uncertainty   &         &                      & 0.0122 & 0.0035 \\
\enddata
\end{deluxetable*}


\section{Astrometric Residuals and Limits on Third Bodies\label{sec:3rdbody}}

\begin{figure*}[ht]
\centering
\includegraphics[width=3.9in]{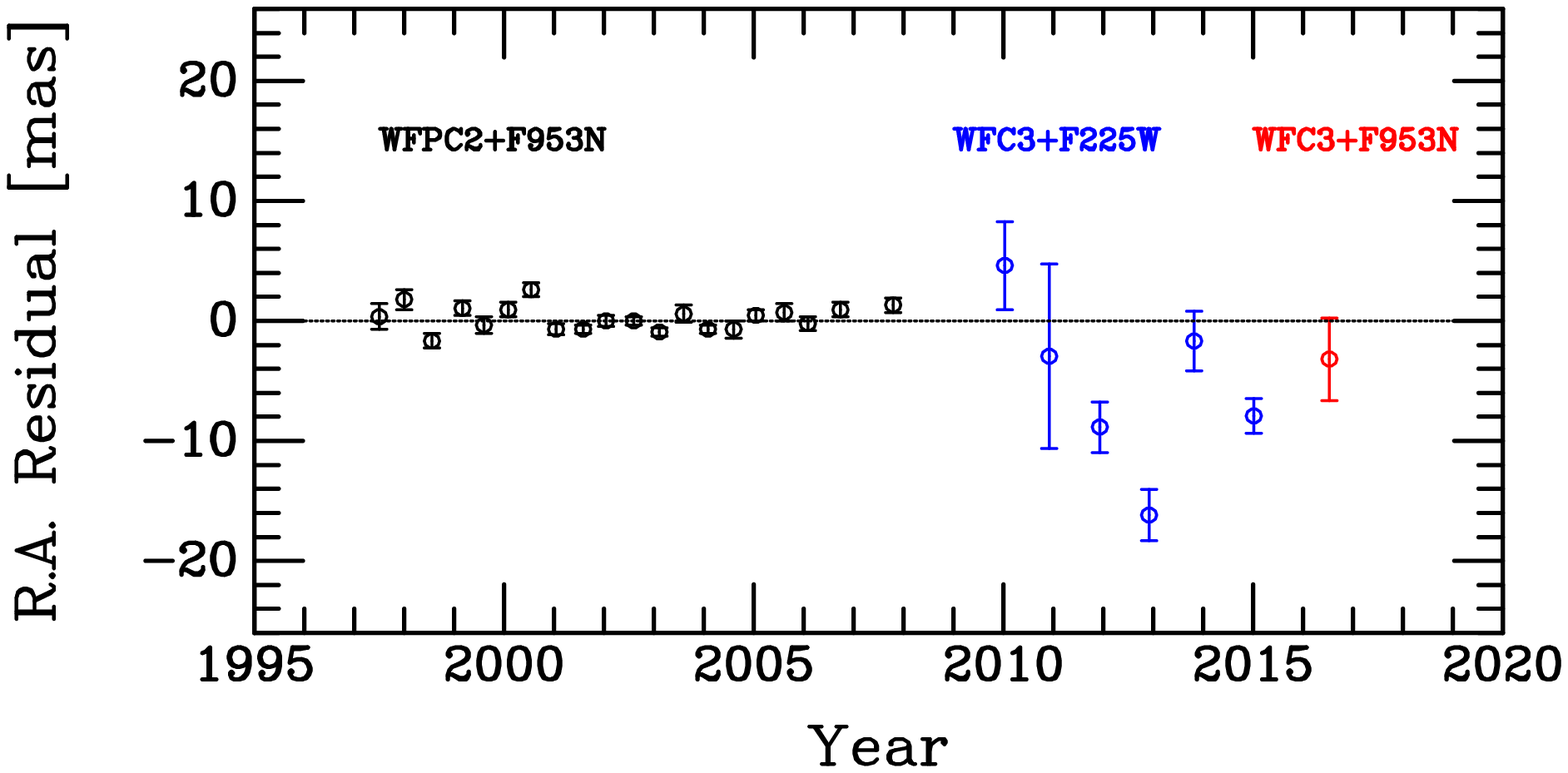}
\vskip0.25in
\includegraphics[width=3.9in]{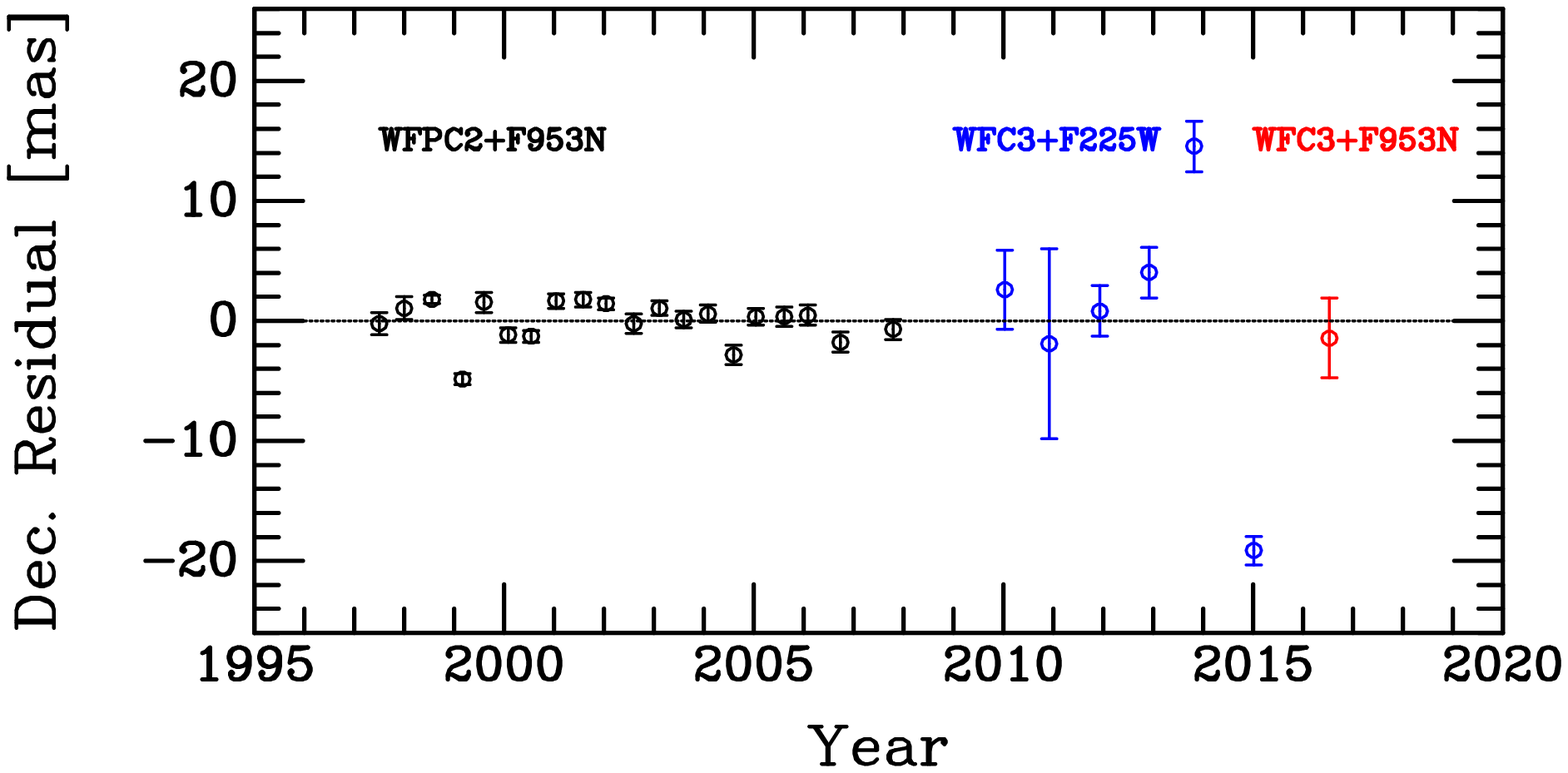}
\caption{Residuals (in the sense observed minus ephemeris) for the \HST\/ astrometry, in right ascension ({\it top panel}) and declination ({\it bottom panel}). Color coding indicates the three different camera-plus-filter combinations that were used. 
\label{fig:resids}
}
\end{figure*}

In Figure~\ref{fig:resids} we plot the residuals between the \HST\/ astrometric measurements and our final orbital solution (\S\ref{sec:solution} and Table~\ref{table:elements}), versus observation date. The top panel shows the residuals in right ascension, and the bottom panel plots them in declination. Black points represent the residuals for the WFPC2+F953N observations; blue shows them for WFC3+F225W; and red plots them for the single WFC3+F953N measurement. The WFC3 data have relatively large error bars and a few very large residuals; these are plausibly due to uncertainties in the correction for chromatic aberration for the F225W data, and also to the large magnitude difference in the WFC3 observations, an increasingly important source of error as the binary separation decreased. The residuals for the WFPC2+F953N combination are much smaller, but still have some outliers that exceed the formal uncertainties. These likely arise from telescope breathing and CTI, as discussed above. 

Figure~\ref{fig:resids} indicates that there is no
convincing evidence for periodic perturbations with semi-amplitudes of more than
$\sim$2--3~mas, based on the WFPC2+F953N astrometry. The long-term stability of third bodies orbiting around individual stars in a binary
system has been studied numerically by, among others, \citet{Holman1999}.
Using the results in their Table~3, and the parameters of the present-day \mucas\
binary, we find that the longest periods for stable third-body orbits in the
system are about 1.07~yr for a body orbiting \mucas~A, and 0.74~yr for
one orbiting \mucas~B\null. 

\begin{figure*}[ht]
\centering
\includegraphics[width=3.9in]{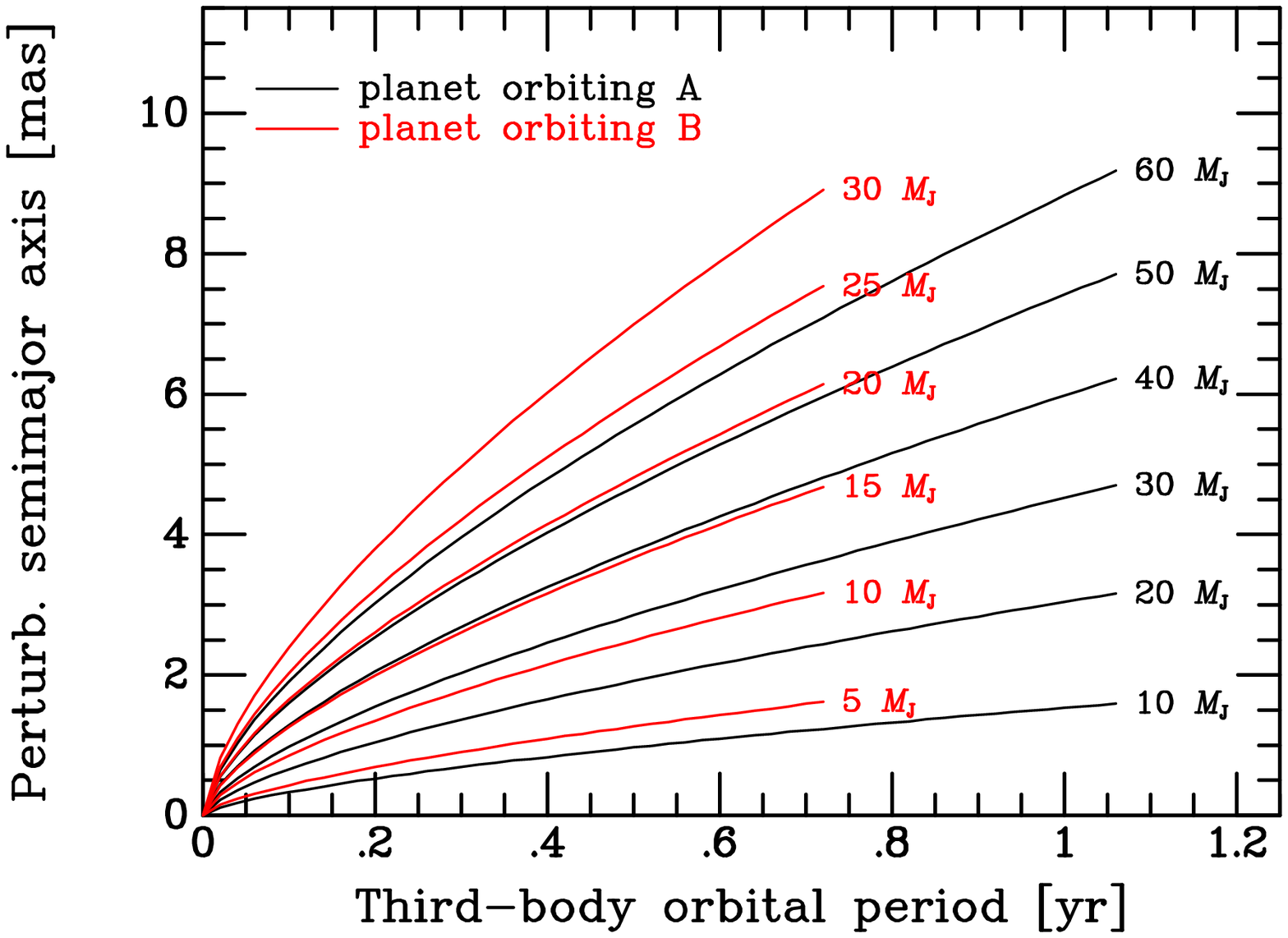} 
\caption{
Astrometric perturbations that would result from planetary or brown-dwarf companions of
\mucas~A ({\it black curves}) or \mucas~B ({\it red curves}), with the masses
of the perturbers (in units of the Jovian mass) indicated in the labels.
Calculations were made for periods up to the orbital-stability limits of bodies
with orbital periods of $\sim$1.07~yr (companions of \mucas~A) or $\sim$0.74~yr
(companions of \mucas~B). The $y$-axis is the semimajor axis of the resulting
astrometric perturbation of A or B in milliarcseconds.
\label{fig:perturbs}
}
\end{figure*}

We calculated the semimajor axes of the astrometric perturbations of both stars
that would result from being orbited by substellar companions of masses ranging
from 5 to $60\,\Mjup$ (where $\Mjup$ is the mass of Jupiter,
$0.000955\,M_\odot$), and for orbital periods up to the stability limits given
above. The results are plotted in Figure~\ref{fig:perturbs}. For a semi-amplitude limit of 3~mas, the figure indicates that companions of \mucas~A or B of $\sim\!20\,\Mjup$ or less, or $\sim\!8\,\Mjup$ or less, respectively,
could escape astrometric detection at periods close to the stability limit. At shorter periods, successively higher masses could go undetected by astrometry. High-precision RV studies of \mucas~A could set tighter limits on third bodies orbiting the primary star at short periods.

\section{Astrophysical Parameters of \mucas~A}

As a well-known bright, moderately metal-poor, high-velocity star, \mucas~A has been the subject of numerous observational investigations. It is included among three dozen well-studied ``\Gaia\/ FGK benchmark stars'' \citep[e.g.,][and references therein]{Jofre2014, Heiter2015, Jofre2018}. Observational data on \mucas~A were reviewed extensively five years ago by \citet{Bach2015}. We discuss and update the astrophysical parameters of the star in this section.

\subsection{Angular Diameter \label{sec:angdiam}}

The angular diameter of \mucas~A was measured with the CHARA Array interferometer and reported by \citet[][hereafter B08]{Boyajian2008}. They obtained a physical diameter (corrected for limb darkening) of $0.973\pm0.009$~mas. At the distance of \mucas\ (Table~\ref{table:parallax}), this corresponds to a physical radius of $0.789\pm0.008\,R_\odot$. When the distance and absolute luminosity of a star are known, measurement of the angular diameter allows its effective temperature, $\Teff$, to be calculated from first principles. This can be compared with $\Teff$ values inferred from spectroscopic and\slash or photometric data.



A few recent authors have noted discrepancies between effective temperatures of stars determined from spectroscopic and photometric measurements and those obtained from interferometric diameters \citep[e.g.,][]{Casagrande2014, Karovicova2018, White2018}. These discordances typically arise for stars with angular diameters close to the interferometric resolution limit ($\la$1~mas) in the near-infrared $K$ band, in the sense that the measured sizes often appear systematically larger than expected from the spectroscopic or photometric effective temperatures. To investigate whether the B08 diameter measurement of \mucas~A could have been impacted by these possible systematics, we re-reduced the archival B08 data by passing them through the most recent version of the reduction pipeline for CHARA Classic data.\footnote{\url{http://www.chara.gsu.edu/tutorials/classic-data-reduction}}  

The latest code differs in the method used to compute the visibilities (integrating the power spectrum, versus fringe-fitting functions), and the computation of the noise.  Our new reduction yields a diameter 8\% smaller, and thus an effective temperature 4\% larger, than the values derived by B08. The difference between the old and new results suggests that there could be a problem with the interferometric diameter of \mucas~A published by B08. However, a full evaluation of this apparent discrepancy is beyond the scope of the present paper. The B08 measurement was made in the $K$ band in the near-infrared. We note that a measurement of the angular diameter of \mucas~A using higher-spatial-resolution observations in the near-infrared $H$ band, or at visible wavelengths, could provide a tighter constraint. In the meantime, in the following discussion we will retain the B08 diameter measurement.

\subsection{Chemical Composition \label{sec:composition}}

For the \Gaia\/ benchmark stars, there is detailed information available on their chemical compositions, assembled from extensive high-resolution spectroscopic studies. \citet{Jofre2018} tabulate abundances of 20 individual metals in \mucas~A\null. Further details are given by \citet{Casamiquela2020}. The carbon and oxygen contents of the star have been determined by \citet{Luck2005} and \citet{Luck2017}. 


\subsection{Stellar Parameters} 

In Table~\ref{table:parameters} we list physical parameters of \mucas~A and the literature sources from which they are quoted. Row one gives the dynamical mass determined in this paper. The radius and absolute luminosity in rows two and three of the table are corrected slightly from the cited literature values by adopting the parallax we give in Table~\ref{table:parallax}.

Since we have directly measured the mass of \mucas~A, and its radius and luminosity are known, we can calculate its effective temperature and surface gravity from first principles. These are listed in rows four and five of the table. We find $\Teff=5306\pm31$~K\null. A compilation of the parameters $\Teff$ and $\log g$ for \mucas~A from published spectroscopic and photometric analyses is given by \citet{Heiter2015}. For the effective temperature, they find a mean of $\Teff=5341$~K, with an appreciable standard deviation of 92~K, from seven determinations based on spectroscopy; and 5338~K with $\sigma=82$~K from four photometric determinations. These values are slightly ($\sim$1$\sigma$) higher than the $\Teff$ that we calculate directly from the radius and luminosity. An even higher effective temperature of 5403~K\footnote{An uncertainty was not given explicitly, but is probably about $\pm$50~K, judging from their findings for the majority of the stars in their sample.} was found by \citet{crm10}, based on the infrared-flux method (IRFM)\null. The more recent PASTEL literature compilation \citep[][version of 2020 January 30]{Soubiran2016}\footnote{\url{https://vizier.u-strasbg.fr/viz-bin/VizieR?-source=B/pastel}} lists some 39 determinations of the atmospheric parameters of \mucas~A (although some are re-publications of the same values). Among the 26 published in the 21st century, the effective temperatures range from 5240 to 5720~K\null.  We return to the subject of the effective temperature in the next section. 

For the surface gravity, eight determinations from spectroscopic analyses, summarized by \citet{Heiter2015}, gave a mean of $\log g=4.51$ with $\sigma=0.20$. Here the agreement with our value of $4.515\pm0.011$, calculated from the radius and the dynamical mass, is excellent. 


Rows six through nine of Table~\ref{table:parameters} summarize the chemical composition (see \S\ref{sec:composition}). Row ten gives the rotational velocity, $v\sin i$. The final two rows of the table give the $V$ magnitude and $B-V$ color of the \mucas\ system, from the literature compilation by \citet{me91}.

As noted by B08, the effective temperature, luminosity, and radius of \mucas~A are approximately those of a normal Population~I K0~V dwarf; and we now see that this is also true of the mass. The fact that the star had been considered to be a standard star with an earlier spectral type of G5~V \citep[e.g.,][]{Keenan1953} is a result of the metallic-line weakening caused by the star's metal deficiency, $\rm[Fe/H]=-0.81$. 

Less than half a dozen field late-type dwarfs as metal-deficient as \mucas~A have had dynamical-mass determinations \citep[e.g.,][]{Jao2016}. Of these, the mass of \mucas~A now has by far the highest precision. Precise masses and radii have also been derived for the components of several eclipsing binaries in metal-poor globular clusters (GCs) \citep[][and references therein]{Thompson2020}.


\begin{deluxetable}{lcc}
\tablewidth{0 pt}
\tablecaption{Physical Properties of \mucas~A 
\label{table:parameters}
}
\tablehead{
\colhead{Parameter} &
\colhead{Value} &
\colhead{Source\tablenotemark{a}}
}
\startdata
Mass, $M$          & $0.7440\pm0.0122\,\,M_\odot$     & (1) \\
Radius, $R$        & $0.789\pm0.008\,\,R_\odot$     & (2) \\
Luminosity, $L$    & $0.445\pm0.005\,\,L_\odot$     & (3) \\
Effective temperature, $\Teff$ & $5306\pm 31$ K           & (4) \\
Surface gravity, $\log g$ [cgs] & $4.515\pm 0.011$        & (5) \\
Surface iron abundance, [Fe/H]        & $-0.81\pm0.03$                          & (6) \\
$\alpha$-element abundance, $[\rm\alpha/Fe]$ & +0.3 & (7) \\
Carbon abundance, [C/Fe]         & $+0.08\pm0.08$                          & (8) \\
Oxygen abundance, [O/Fe]         & $+0.56\pm0.08$                          & (8) \\
Rotational velocity, $v\sin i$  & $2.4\,\kms$                             & (8) \\
$V$                             & $5.166\pm0.014$                         & (9) \\
$B-V$                           & $0.695\pm0.006$                         & (9) \\
\enddata
\tablenotetext{a}{Sources: (1) This paper, Table~\ref{table:masses}; (2) Boyajian et al.\ 2008, adjusted for parallax;  (3) Heiter et al.\ 2015, adjusted for parallax; (4) This paper, calculated from $L$ and $R$, and assuming a solar $\Teff=5771\pm1$~K from Heiter et al.; (5) This paper, calculated from $M$ and $R$; (6) Jofr\'e et al.\ 2014, 2018; (7) mean of Mg, Si, Ca, and Ti, from Jofr\'e et al.\ 2018; (8) Luck 2017, C and O abundances converted from his [C/H] and [O/H] values using his $\rm[Fe/H]=-0.75$; (9) Combined light of AB system, from Mermilliod 1991. 
} 
\end{deluxetable}







\newcommand\pp{{\phantom{$+$}}}
\newcommand\msol{{\cal M_{\odot}}}
\newcommand\teff{{T_{\rm eff}}}
\newcommand\lta{\mathrel{\hbox{\raise 0.6 ex \hbox{$<$}\kern
                   -1.8 ex\lower .5 ex\hbox{$\sim$}}}}
\newcommand\gta{\mathrel{\hbox{\raise 0.6 ex \hbox{$>$}\kern
                   -1.7 ex\lower .5 ex\hbox{$\sim$}}}}





\section{Astrophysical Implications}
\label{sec:astrophysics}

As outlined in the previous section, there remain uncertainties in the effective temperature and other parameters of \mucas~A\null. Moreover, we have raised a possible concern about the measurement of its angular diameter (\S\ref{sec:angdiam}). Effective temperatures near $\Teff=5340~K$ have been found in many spectroscopic
and photometric studies, but somewhat higher temperatures are generally found from application of the IRFM\null. A significant increase in the adopted temperature of $\mu\,$Cas~A would certainly impact, e.g., the metallicity of $\rm[Fe/H] = -0.81$ (see Table~\ref{table:parameters}), which was derived assuming $\teff = 5308$~K and LTE conditions.  As a 
rule of thumb, increasing the temperature by 100~K will result in about a 0.1~dex
increase in the [Fe/H] value derived from \ion{Fe}{1} lines---although
metallicities based on \ion{Fe}{2} lines are known to be much less sensitive to
$\teff$.  Fortunately, corrections for non-LTE effects appear to be small for
stars with intrinsic properties (temperatures, gravities, and metallicities)
similar to those of $\mu\,$Cas~A ($\lta$0.02 dex, according to \citealt{lba12}; see their Figure~2).

Since the effective temperature and diameter of $\mu\,$Cas~A remain
uncertain, it is not possible to constrain its age and helium abundance as tightly as we had hoped---inspired by D65---when we began our {\it HST\/} program. Nevertheless, with reasonable assumptions for its
$\teff$\ and metallicity, together with the considerably improved mass precision that is the main
result of our study, comparisons of theoretical predictions for the mass-radius 
and mass-luminosity diagrams with the observed quantities can be made. These should provide some
indication whether $\mu\,$Cas~A has close to the primordial helium
abundance ($Y_p \approx 0.247$; see \citealt{cfo16}). We should also be able to make an improved estimate of the star's age. Previous age determinations for \mucas~A have varied remarkably widely, from about 2--3 to 11 Gyr (e.g., \citealt{nma04}; \citealt{mh08}; \citealt{Luck2017}).

Rather than adopting, say, the mean $\teff$ and [Fe/H] values from recent
publications, which may or may not be entirely consistent with each other, we
decided to rely on photometric observations of $\mu\,$Cas~A, subject to
constraints provided by the GC 47~Tucanae. We will also take into account the
interferometric angular diameter (B08), our updated parallax (Table~\ref{table:parallax}), and our new precise mass
determination (Table~\ref{table:masses}).  

The main role of 47 Tuc in our discussion below is to calibrate the
transformations between $B-V$ and $\teff$, and thereby tie our results for the
field metal-poor star $\mu\,$Cas~A to its counterparts in a GC with nearly
the same metallicity, and presumably a similar age.  We were motivated to
do this because we noticed that $\mu\,$Cas~A is intrinsically redder than
main-sequence (MS) stars in 47 Tuc at the same absolute magnitude, which is the
opposite of what is expected if the star is slightly more metal-poor than the cluster, as indicated by most [Fe/H] determinations.
Importantly, this approach ensures that the properties of $\mu\,$Cas~A are derived
in a fully consistent way, although the accuracy of our findings in an absolute
sense will depend on various factors, including, in particular, the assumed
radius and adopted $B-V$ color of $\mu\,$Cas~A and the photometry, metallicity,
and reddening of 47 Tuc.  Consistent with the $E(B-V)$ value that is obtained
from the three-dimensional extinction maps of
\citet{clv17},\footnote{\url{http://stilism.obspm.fr}} we will assume that
$\mu\,$Cas is unreddened.

\subsection{Constraints from 47 Tucanae}
\label{subsec:47tuc}

\begin{figure*}[ht]
\plotone{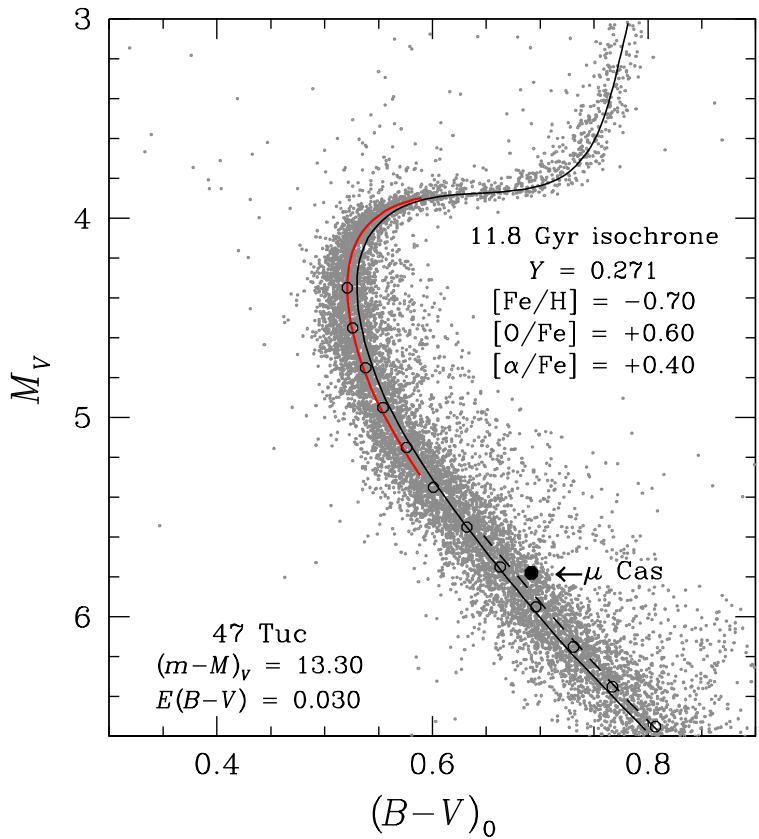}
\caption{Comparison of the locations of $\mu\,$Cas~A and main-sequence (MS) stars
in 47 Tucanae in the [$(B-V)_0,\,M_V$]-diagram.  Plotted as open circles are median
points that were derived from the binning of the cluster stars in intervals of
0.2 mag in $V$.  The solid curve in black represents an 11.8 Gyr isochrone for
the indicated chemical abundances, where [$\alpha$/Fe] includes all of the
$\alpha$ elements except oxygen, which is specified explicitly.
If this isochrone is adjusted by $\delta(B-V) = 0.009$ mag to the blue, it
reproduces the observed turnoff (TO) luminosity, as shown by the red curve.
(For clarity, only the turnoff portion of the latter is plotted.)
The dashed curve shows where the lower MS of an otherwise identical isochrone
as the solid black curve, but for the helium content lowered to $Y = 0.25$, would be located.}
\label{fig:f1}
\end{figure*}

Figure~\ref{fig:f1} shows where $\mu\,$Cas~A is located relative to MS stars in the color-magnitude diagram (CMD) of 47 Tuc, which is assumed to have the apparent distance modulus and reddening
that are specified in the lower left-hand corner of the plot.  (We have used
the cluster photometry that is publicly available in P.~B.~Stetson's
``Homogeneous Photometry"
archive;\footnote{\url{www.cadc.hia.nrc.gc.ca/en/community/STETSON/homogeneous}}
these observations are discussed by \citealt{bs09}.)  \citet{bvb17}
studied the available determinations of the foreground 
reddening of the cluster, and concluded that $E(B-V) = 0.03 \pm 0.01$ is the current best
estimate. This value differs from the mean values derived from dust maps
(\citealt{sfd98}; \citealt{sf11}; \citealt{clv17}) by only 0.002--0.004~mag.  

The distance moduli for 47~Tuc
that have been derived in recent studies, using independent methods,
are also in superb agreement.  Brogaard et al.~used the eclipsing-binary member
V69 to obtain $(m-M)_V = 13.30\pm 0.06$, which happens to be midway
between the values of 13.27 from simulations of the horizontal-branch (HB)
population of 47 Tuc \citep{dvk17}, and 13.33 from {\it Gaia\/} DR2 parallaxes 
\citep{crc18}, both of which have similar uncertainties.  The latter
estimate, which used background stars in
the Small Magellanic Cloud and quasars to account for 
spatial and magnitude variations of the parallax zero point (see 
\citealt{lhb18}), is particularly noteworthy because it is model independent.

We determined the position of $\mu\,$Cas~A plotted in Figure~\ref{fig:f1} as follows.
According to the compilation by \citet{me91}, which tabulates homogeneous
mean photometry in the $UBV$ system, $\mu\,$Cas has $V = 5.166 \pm 0.014$ and
$B-V = 0.695 \pm 0.006$.  Its distance of 7.54 pc, calculated from our
adopted parallax (Table~\ref{table:parallax}), corresponds to a true distance modulus of
$(m-M)_0 = -0.614$. From this we obtain $M_V = +5.78$. The photometric properties of $\mu\,$Cas~A itself
will be slightly different because of the presence of the low-mass companion (see
Table~\ref{table:masses}).  Its contribution can be estimated by 
determining its absolute magnitudes in the $B$ and
$V$ bandpasses from stellar models for its measured mass and the metallicity of the system.
Subtracting these contributions from the observed luminosities, we find
$V \simeq 5.170$ and $B-V \simeq 0.692$ for \mucas~A\null.  Even
though the bolometric corrections (BCs) applicable to very low-mass stars have
significant uncertainties, the effects of the companion on the observed
photometry amount to no more than a few thousandths of a magnitude.

Superimposed on the CMD in Figure~\ref{fig:f1} is an 11.8~Gyr isochrone for
the indicated chemical abundances.  It has been transposed from the theoretical
to the observed plane using the color--$\teff$\ relations given by
\citet{cv14}. This is the same isochrone used by \citet{bvb17} to fit
both the turnoff (TO) photometry of 47 Tuc and the properties of the eclipsing
binary V69.  Were it not for the binary, it would be difficult to argue
against the possibility that 47 Tuc has a lower metallicity
or reduced abundances of oxygen and/or the other $\alpha$\ elements.  In 
addition, according to \citet{dvk17}, the distribution of HB stars in 47 Tuc
can be reproduced very well by synthetic HBs if there is a star-to-star
variation of the initial He abundance by $\Delta(Y) \simeq 0.03$, with a mean
abundance corresponding to $\langle Y\rangle \simeq 0.271$.  Cluster MS
stars with this He abundance are therefore assumed to lie along the median
fiducial sequence that has been plotted as open 
circles.\footnote{As discussed by \citet{vbl13}, the best estimate of the TO
age is obtained by first shifting in the horizontal direction all of the
isochrones for a suitable range in age until each of them matches the observed
TO color, and then determining which one provides the best superposition of the
subgiant stars located just past the TO\null.  Figure~\ref{fig:f1} shows that, if the
isochrone in black were adjusted to the location of the red curve, it would
provide a good fit to the TO and subgiant stars.  The advantage of this
procedure is that errors in predicted or observed colors have little or no
impact on the derived age.}  The dashed curve, which represents the lower MS
portion of an isochrone for the same age and metal abundances, but with the helium content reduced to
$Y = 0.250$, serves to illustrate the dependence of predicted MS loci on~$Y$.

There are systematic differences between the isochrone and the median
cluster fiducial, in the sense that the models are too red by about 0.01 mag
in the vicinity of the TO and by a similar amount, but in the opposite sense, at
$M_V = +6.6$. However, the predicted $B-V$ colors at the absolute magnitude of $\mu\,$Cas~A
are too blue by only 0.003 mag.  Although this offset is quite small, it should
(and will) be taken into account when we fit isochrones to the photometric
properties of $\mu\,$Cas~A\null.  Thus, we have learned from our study of 47 Tuc 
that the $B-V$ colors that are derived from the tables of BCs given by
\citet{cv14} should be adjusted to the red by 0.003 mag
when applied to stars with properties similar to those of $\mu\,$Cas~A\null.  To be
sure, many assumptions have been made in reaching this point, such as the 
reddening and the helium and metal abundances of 47 Tuc, but the best estimates
of the various parameters result in $\mu\,$Cas~A lying slightly to the red of
the cluster MS\null. The goal of our analysis now is to understand why that is, and
what the implications are for the properties of $\mu\,$Cas~A.

Having established the absolute $V$ magnitude of $\mu\,$Cas~A and its $B-V$
color, we can use our tables of BCs to convert $M_V$ to $M_{\rm bol}$, and thence to
the bolometric luminosity.  The BC tables require input values of $\teff$ and
[Fe/H].  (They also depend on [$\alpha$/Fe], but we have opted to assume that
[$\alpha$/Fe] $= +0.3$, given the support for this value from observational
studies; see Table~\ref{table:parameters}.)  Since the luminosity can
also be calculated from the radius and $\teff$, it is necessary to iterate on
the input parameters until (1)~both ways of calculating the luminosity yield the
same result, and (2)~the predicted $B-V$ color (including the small offset
described above) matches the observed color.  With just a few iterations of this 
procedure, we obtained [Fe/H] $= - 0.74$, $\teff = 5346$~K, and $L/L_\odot =
0.458$.  (We have assumed that this temperature has an uncertainty of $\pm 70$~K,
mainly so that the error bar encompasses the effective temperatures derived from both
interferometric studies and the IRFM\null.  For the luminosity uncertainty, we have
adopted $\pm$0.014$\,L_\odot$,
which corresponds to a 0.03--0.035
mag uncertainty of our bolometric corrections.)
Remarkably, this temperature is within a few Kelvin of the mean
spectroscopic and photometric determinations tabulated by \citet{Heiter2015}, and
the metallicity is within 0.07 dex of the values given by Heiter et al.\ and
\citet{Luck2017}.
Our determination of the luminosity is within 1$\sigma$ of the value of
$L/L_\odot = 0.445 \pm 0.005$ given by Heiter et al.\ (our Table~\ref{table:parameters}).
As we have adopted B08's determination of the diameter of
$\mu\,$Cas~A in our analysis, but found a higher temperature by $\approx 50$~K,
the bolometric flux that B08 derived must be smaller than our estimate.


The problem remains that the photospheric metallicity of $\mu\,$Cas~A appears to be lower than
that of 47~Tuc, and yet it is redder than cluster stars of the same
absolute magnitude.  This is very likely the consequence of diffusive
processes.  It is now well established that diffusion acts to reduce the
abundances of He and the metals in the surface layers of old stars, although
extra mixing below surface convection zones must also be present to limit the
efficiency of gravitational settling.  Otherwise, as shown by \citet{rmr02},
diffusive models would be unable to explain the observed variation of the Li
abundance with $\teff$\ in the so-called ``Spite-plateau" stars
(\citealt{ss82}) or the abundance variations between the TO and lower red-giant
branch (RGB) in GCs.  

In the lowest-metallicity GCs, such as NGC\,6397,
observations have revealed that the difference in metallicity between the TO and
lower RGB is about 0.15 dex (\citealt{kgr07}; \citealt{nkr12}), which is in
rather good agreement with the expectations from diffusive stellar models
with extra mixing (see \citealt[their Figure~9, concerning the very metal-deficient
cluster M92]{vrm02}).  At intermediate metallicities, the variation in
[Fe/H] appears to be somewhat less; e.g., \citet{gnk14} have found a difference
of about 0.1 dex across the subgiant branch.  \citet{mmc16} found an even
smaller difference in 47 Tuc, although the uncertainties are such that the
variation could be anywhere in the range from 0.0 to 0.1 dex.  Part of the
difficulty is that the difference in [Fe/H] between the TO and RGB is quite
dependent on the assumed $\teff$\ scale (see \citealt[their Table 5]{gnk14}).
Regardless, the signature of diffusion appears to have been detected in the
near solar-abundance open cluster M$\,$67 as well (\citealt{ogk14};
\citealt{bpr18}), but its effects are much smaller ($\Delta$[Fe/H] $\simeq
0.04$--0.05\ dex), probably due mostly to its considerably younger age.

In any case, it is reasonable to assume that [Fe/H] values for $\mu\,$Cas~A
that are derived from spectroscopic or photometric studies should be increased
by about 0.08 dex in order to obtain the metallicity that applies to its
interior structure.  Therefore, the isochrones that are used to interpret the
observations of this star should assume [Fe/H] $\approx -0.66$.  Indeed, it is
this value of [Fe/H] that should be compared with the metallicity of 47 Tuc
that has been inferred from the binary V69, because that is the relevant metal
abundance for the calculation of the mass-radius and mass-luminosity relations 
that are used in comparisons with the measured masses and radii and the
derived luminosities of the components of the binary.  By the same token, the surface
metallicities of upper-MS stars in 47 Tuc should be less than that of $\mu\,$Cas~A\null.
Clearly, a small difference in [Fe/H], with 47 Tuc being more metal-poor than
$\mu\,$Cas, would help to explain the small offset of the latter relative to
cluster MS stars of the same $M_V$ on the [$(B-V)_0,\,M_V$]-diagram.  A
difference in the He abundance may also be partly responsible for the
difference in color (note in Figure~\ref{fig:f1} the separation of the solid and
dashed curves in the vicinity of $\mu\,$Cas~A).\footnote{As noted by the referee, the formation and early evolution of a
binary star occurs in a very different environment than in the case
of an isolated, single star.  It is possible that this could give rise to small differences in
their CMD properties at the same mass, age, and chemical composition.}

In principle, spectroscopy of cluster giants should yield
a metallicity that is close to the initial metal abundance, because deepening
convection along the lower RGB will dredge back into the surface layers
most (but not all) of the helium and the metals that had settled into the
interior during the core H-burning phase; i.e., the effects of diffusion are
mostly erased during the evolution along the lower RGB\null.  However, due mainly to
systematic uncertainties, it is difficult to derive absolute metal abundances to
within $\sim$0.1 dex.  Although, for instance, the new metallicity scale
developed by \citet{cbg09} gives [Fe/H] $= -0.76$ for 47 Tuc, higher or lower
values by 0.05 to 0.1 dex are commonly found (e.g., \citealt{cpj14};
\citealt{wpc17}; \citealt{tsa18}).

\subsection{The Age and Helium Abundance of $\mu\,$Cas}
\label{subsec:age}

\begin{figure}
\plotone{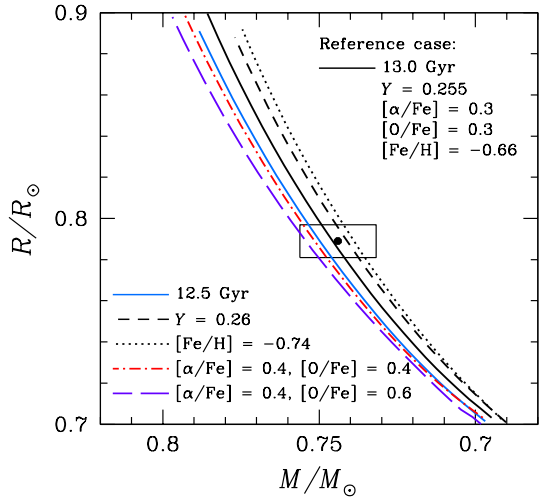}
\caption{Comparison of predicted mass-radius relations from isochrones for the
indicated ages and chemical abundances with the mass and radius of $\mu\,$Cas~A
(filled circle and error box).  The solid curve in black represents the
reference case; the others illustrate the effects of varying the age, $Y$, and
the metal abundances, in turn, as indicated in the lower left-hand corner.  The 
various abundance choices correspond to initial chemical compositions.  Due to
the operation of diffusive and extra mixing processes since $\mu\,$Cas formed,
it is expected that its surface metallicity and He abundance will have changed
by $\Delta$[Fe/H] $\approx -0.08$ dex and $\Delta\,Y \approx -0.03$,
respectively (see the text).  Note that colors are not affected by the modest
variations in the atmospheric abundance of helium that are predicted to occur
during MS evolution.}
\label{fig:zmucasr}
\end{figure}

Using our adopted or derived properties for $\mu\,$Cas~A, we can now consider
their implications for its age and helium content.  Figure~\ref{fig:zmucasr} shows
the observed mass and radius of the star and the associated error bars (the filled circle
and error box), and superposes the predicted mass-radius relations from several isochrones.
The latter were generated from the grids provided by \citet{vbf14}, with the
exception of a set of models that was computed for  [O/Fe] $= +0.6$, using the
same stellar evolutionary code.  Unfortunately, we can produce models only
for [$\alpha$/Fe] $=$ [O/Fe] $= +0.3$ and +0.4, and for [O/Fe] $= +0.6$ when
[$\alpha$/Fe] $= +0.4$ is adopted for the other $\alpha$ elements (due
to the lack of low-temperature opacities for other mixtures of the metal
abundances).  The solid curve, as defined in the legend at the top, right-hand corner of the figure
represents the adopted reference case,
while the others (see legend at the bottom, left-hand corner) assume all of the same
parameter values except for the changes that are given explicitly.
The figure shows that models for high ages and low values of $Y$ provide good fits
to the observed properties of \mucas~A\null. This is not surprising, given that a metal-poor,
high-velocity star in the solar neighborhood is likely to have formed early in
the evolution of the Milky Way. 

The solid blue curve in Figure~\ref{fig:zmucasr} shows the effect of reducing the age of the star by 0.5~Gyr. Lower
ages imply higher masses; i.e., the predicted $M$--$R$ relation is shifted to the
left.  The differences between the reference curve and the dashed locus
illustrate the consequences of increasing the He abundance by $\Delta\,Y =
0.005$; higher $Y$ results in a lower mass at a fixed radius.  A similar, but
somewhat larger, offset (in the same direction) is obtained if the adopted
[Fe/H] value is reduced by 0.08 dex; compare the location of the
reference curve with that of the dotted curve.  In addition, the location of
the dot-dashed locus relative to the reference case shows the effect of
increasing the $\alpha$-element abundances from [$\alpha$/Fe] $= +0.3$ to
$+0.4$.  In this case, the predicted $M$--$R$ relation is shifted to higher
masses, which also occurs if the assumed oxygen abundance is increased by 0.2
dex; compare the dot-dashed and long-dashed curves.

An age of 13.0 Gyr was assumed for most of the isochrones, because this estimate
is close to the maximum possible age of $\mu\,$Cas, given that the Big Bang
apparently occurred about 13.8 Gyr ago (\citealt{blw13}; \citealt{pla14}).
Although the predicted mass-radius
relation that applies to $\mu\,$Cas~A could be slightly to the right of those that
have been plotted---which is the direction of increasing age---it is perhaps
somewhat more probable that the relevant relation is located somewhat to the
left.  The permitted age range therefore depends on how far the predicted
$M$--$R$ relation is from the lower left-hand corner of the
error box.  For instance, for the cases plotted as dot-dashed and long-dashed
curves, the mass-radius relations for ages less than { 11.5--12.0}~Gyr would
lie outside the error box. 

Obviously, the uncertainties can accommodate much
larger ranges in age if a higher He abundance or a lower metallicity is assumed,
since these cases (the dashed and dotted loci) are located further to the right
than any of the others.  In fact, although the isochrones were generated for He
abundances in the range $0.25 \le Y \le 0.26$, there is ample room within the
error box that, e.g., the dot-dashed and long-dashed cases could be made to
satisfy the observational constraints for ages well below { 11}~Gyr, provided that
the assumed He abundance is increased by a sufficient amount.  According to
Figure~\ref{fig:zmucasr}, an increase in $Y$ by 0.005 has almost the same effect on
the mass-radius relation as an increase in age by 0.5 Gyr; that is, such changes, if
made simultaneously { to the models}, result in essentially the same relation between mass and
radius.  Thus, the current uncertainties associated with the
mass and radius of $\mu\,$Cas~A can potentially accommodate a fairly large range
in $Y$ and age (although it would be very surprising if this star has $Y > 0.26$).

\begin{figure*}[ht]
\plotone{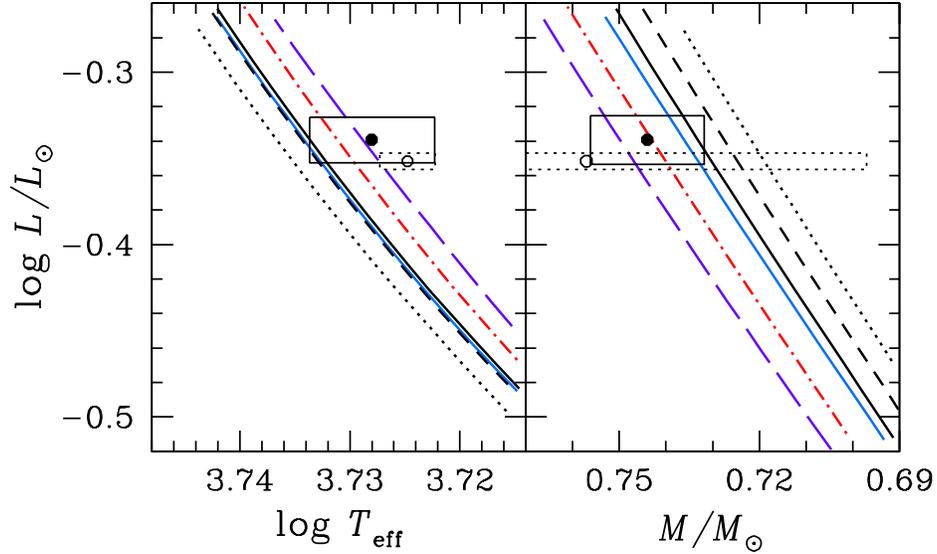}
\caption{Isochrones from the previous figure, with the same color and line-type encoding, except that comparisons with the properties of \mucas~A are made on the H-R diagram
(left-hand panel) and the mass-luminosity diagram (right-hand panel). The filled circles and solid error boxes show our measured and adopted values. The open circles and dotted error boxes represent the $\teff$ and luminosity that
were derived by \citet{Boyajian2008} and \citet{Heiter2015} respectively (but corrected for our adopted parallax), along with the best estimate of the mass of \mucas~A before the present study, $0.757 \pm 0.060\, M_\odot$ \citep[from][]{Lebreton1999}.}
\label{fig:zmucas}
\end{figure*}

Figure~\ref{fig:zmucas} compares the same isochrones with the observed properties
of $\mu\,$Cas~A on the H-R diagram and the mass-luminosity plane. In these two panels, we plot \mucas~A at $L=0.458\pm0.014\,L_\odot$ and $\Teff=5346\pm70$~K from the discussion in \S\ref{subsec:47tuc}, and at the dynamical mass of $0.7440\pm0.0122\,M_\odot$ determined in this paper.  
We use filled circles and solid lines for the associated error boxes for these results. To compare with earlier results, we use open circles and dotted error boxes to plot the location of the star based on the $\teff$ from B08  and the luminosity from \citet{Heiter2015} (both slightly adjusted to take our revised parallax into account), and the mass from Lebreton et al.~(1999), which was acknowledged by B08 to be the best available one at the time.
Interestingly, the dashed locus (which assumed a helium content of $Y=0.26$) is significantly
displaced from the observations in both panels, whereas it provided  { quite a good}  fit
to the observed mass and radius in the previous figure.\footnote{ Note, however, that the assumed metal abundances of these, or any of the other, stellar models do not correspond exactly to the observed abundances; consequently, one cannot determine the ``best-fit" models simply from an inspection of Figures~\ref{fig:zmucasr} and \ref{fig:zmucas}.  The purpose of these two figures is to illustrate how the predicted relations between mass, radius, luminosity, and $\Teff$ are affected when the value of each parameter (age, $Y$, [$\alpha$/Fe], etc.) is individually varied in turn.  Below we show how to use the results of the model computations to derive the overall best estimates of the age and helium content of \mucas~A.}
The discrepancy is even
larger if the dashed curve is compared with the open circles.  However,
predicted $M$--$L$ relations should be more robust than those that involve
temperatures or radii, because the latter are subject to many uncertainties,
including the treatment of convection and the atmospheric boundary condition.
On the other hand, our models generally provide good fits to GC CMDs, especially
to the morphologies of their MS stars (see, e.g., the plots provided by
\citealt{vbf14} and \citealt{cv14}), which suggests that the model
$\teff$\ scale is reasonably good.\footnote{Fits to $BV$ photometry tend
to be somewhat more problematic (recall the discrepancies between theory and
observations in Figure~\ref{fig:f1}) than in the case of $VI_CK_s$ data (also see
\citealt{vcs10}); consequently, the apparent difficulties with $B-V$ colors are
probably due more to deficiencies in the transformations to the $B$ bandpass
than to problems with the predicted effective temperatures.}  

{ Although the dot-dashed curve appears to provide the best consistency between the models and our results for $\mu$~Cas~A (the filled circle) in the mass-luminosity plane, it does not take into account the relatively high abundance of oxygen. In addition, the assumed abundances of the other $\alpha$ elements are slightly too high (see Table~\ref{table:mrl}).}
If the reasonable assumption is made
that oxygen and iron diffuse at a similar rate, the initial and current
[O/Fe] values will be about the same; consequently, the models that are
compared with the observations should assume [O/Fe] $= +0.56$.  We can evaluate
how much the predicted $M$--$L$ relations would be affected by such an oxygen
enhancement, using the data that are listed in Table~\ref{table:mrl}.  This gives
the masses that are obtained by interpolation along the computed $M$--$R$ and
$M$--$L$ relations at, in turn, the observed radius ($0.789\, R_\odot$) and the
luminosity that we have derived ($\log\,L/L_\odot = -0.339$).  The last two
numbers in the right-hand column tell us that $\Delta\rm[O/Fe] = +0.2$ will
increase the predicted mass by $0.0068\,M_\odot$.  Thus, a 0.26-dex increase
(to be in agreement with the observed oxygen abundance) will result in a
mass increased by about $0.0088\, M_\odot$, in which case the mass of our
reference model would increase to $0.7420\, M_\odot$.  This differs from our measured dynamical mass by only $0.0020 \,M_\odot$, which is much less than its uncertainty.

\begin{deluxetable*}{lccccccc}
\tablewidth{0 pt}
\tabletypesize{\footnotesize}
\tablecaption{Predicted Masses for \mucas~A for Various Stellar Models \label{table:mrl}}
\tablehead{\colhead{Stellar Models} & \colhead{Age} & \colhead{$Y$} & \colhead{[Fe/H]} 
 & \colhead{[$\alpha$/Fe]} & \colhead{[O/Fe]} & \colhead{Mass [$M_\odot$]\tablenotemark{a} at}
 & \colhead{Mass [$M_\odot$]\tablenotemark{a} at} \\ 
    \colhead{(Line Type \& Color)\tablenotemark{b}} & \colhead{[Gyr]} & \colhead{ } & \colhead{ } & \colhead{ } 
 & \colhead{ } & \colhead{$R/R_\odot = 0.789$}
 & \colhead{$\log\,L/L_\odot = -0.339$} }
\startdata
\noalign{\smallskip}
Solid (black)\tablenotemark{c} & 13.0 & 0.255 & $-0.66$ & +0.30 & +0.30 & 0.7457 & 0.7332 \\
\noalign{\smallskip}
Solid (blue)  & 12.5 &       &         &      &      & 0.7498 & 0.7362 \\
Dashed (black)       &      & 0.260 &         &      &      & 0.7411 & 0.7276 \\
Dotted (black)       &      &       & $-0.74$ &      &      & 0.7388 & 0.7218 \\
Dot-Dashed (red)   &      &       &         & +0.40 & +0.40 & 0.7514 & 0.7431 \\
Long Dashed (purple)  &      &       &         & +0.40 & +0.60 & 0.7542 & 0.7499 \\
\enddata
\smallskip
\tablenotetext{a}{To be compared with the measured mass of $0.7440\,M_\odot$.}
\tablenotetext{b}{Plotting line type and color used in Figures~\ref{fig:zmucasr} and \ref{fig:zmucas}.}
\tablenotetext{c}{First row is the ``Reference'' case; others assume the same parameter values except as noted.}
\end{deluxetable*}

However, based on the first two entries in the right-hand column of 
Table~\ref{table:mrl}, an { age reduced by 0.5~Gyr will increase} the predicted
mass at the derived luminosity of $\mu\,$Cas~A by $0.0030\, M_\odot$.  Hence the
reference models will predict the observed mass if the assumed oxygen abundance
is increased by 0.26 dex {\it and\/} the age is {decreased to 12.7~Gyr}.
Furthermore, using the tabulated masses for the reference and the dot-dashed
cases, and taking into account the increase in mass by $0.0034\, M_\odot$ if
$\Delta$[O/Fe] $= 0.1$, increased abundances of all of the other $\alpha$
elements, except oxygen, by $\Delta$[$\alpha$/Fe] $= 0.1$ dex apparently
increase the mass by $0.0065\, M_\odot$.  Therefore, if the metal abundances
of the reference model were increased by $\Delta$[$\alpha$/Fe] $= 0.1$ and
$\Delta$[O/Fe] $=0.3$, the predicted mass should increase by $0.0167\,M_\odot$
to $0.7499 \,M_\odot$, which is precisely what the models predict for the
long-dashed case (see Table~\ref{table:mrl}).  Clearly, the tabulated results
are internally self-consistent.

The uncertainties of the derived age and He abundance can also be estimated
using the information that is provided in Table~\ref{table:mrl}.  Because a
reduced age by 0.5 Gyr results in a higher mass by $0.0030\, M_\odot$, the 
$1\,\sigma$ uncertainty in the measured mass, $0.0126\, M_\odot$, is equivalent to an age
uncertainty of { 2.0~Gyr}. In the case of the He abundance: since $\Delta\,Y =
+0.005$ implies a mass reduced by $0.0056\, M_\odot$, the uncertainty of our
mass determination corresponds to a change in $Y$ of 0.011.  However,
because the predicted $M$--$L$ relations are sloped, they will still pass
through the lower left-hand or upper right-hand corners of the error box, if the
mass is lower or higher than the formal uncertainty by $\approx\pm 0.004
M_\odot$; see Figure~\ref{fig:zmucas}.  As a result, the ranges in the age and the 
value of $Y$ that are permitted by the error box are {closer to $\pm$2.7~Gyr and
$\pm$0.014,} respectively.  We therefore conclude that the best estimate of
the age of $\mu\,$Cas { is $12.7\pm 2.7$ Gyr, and that it has a helium content of $Y = 0.255\pm 0.014$}.
(Although we could carry out a similar analysis using the predicted masses
that are derived from the $M$--$R$ diagram, we have not done so because they
would be less trustworthy than those based on the $M$--$L$ diagram, for reasons
already mentioned.) The low rotational velocity (Table~\ref{table:parameters}) is consistent with a high age, but is not decisive.

Unfortunately, our findings remain for now more suggestive than
definitive, because the published interferometric diameter of \mucas\ could
be too large by up to $\sim$8\% (see \S\ref{sec:angdiam}).  Although such a
large correction seems unlikely, given that most spectroscopic and photometric
determinations of $\teff$ (including ours) imply that the measured diameter is
too large by only 1--2\%, it would clearly have important consequences for our
understanding of $\mu\,$Cas if confirmed by future work.  In particular, the
resultant increase in its $\teff$ to $\sim$5500~K would require a re-evaluation
of its metal abundances, perhaps by 0.1--0.15 dex if derived from spectroscopy.
The higher temperature would be especially problematic for photometric
metallicity estimates.  For instance, using either the color--$\teff$\ relations
given by \citet{cv14}, or the semi-empirical transformations provided by
\citet{crm10}, the observed color, $B-V = 0.692$, would imply an [Fe/H] value
$\gta\! -0.4$.
Hopefully, an effort will be made to
obtain a new and improved measurement of the angular diameter of \mucas\ in
the near future, so that it will be possible to resolve this issue.  Needless
to say, any reduction in the uncertainties of its basic properties will help to
improve our understanding of this important star.

\section{Summary and Future Work}

We have obtained high-resolution imaging of the nearby high-velocity visual binary \mucas, using cameras on the {\it Hubble Space Telescope\/} over an interval of nearly two decades. By combining these data with ground-based astrometry of the binary, and with ground- and space-based measurements of the parallax and photocenter motion, we have determined the orbital period (21.568~yr) and calculated dynamical masses for the two components of the binary. We find masses of $0.7440\pm0.0122\,M_\odot$ for the G5~V primary star, \mucas~A, and $0.1728\pm0.0035\,M_\odot$ for its M-dwarf companion, \mucas~B\null. We see no significant indication of perturbations of the \HST\/ astrometric measurements due to a third body in the system.

The main aim of our program was to determine the age and helium content of the moderately metal-poor ($\rm[Fe/H]=-0.81$) primary star \mucas~A\null. However, although we now have a precise dynamical mass for the star, there remain issues with its other astrophysical parameters. We investigated archival interferometric measurements of its angular diameter, leading us to suspect that the diameter may have been overestimated by a few percent. Additionally, there is a fairly wide range of determinations of its effective temperature in the literature. Taking these issues into account, we estimate the age and helium content of the system to be { $12.7\pm2.7$~Gyr and $Y=0.255\pm0.014$.}  
Plotting \mucas~A in the color-magnitude diagram of 47 Tucanae, we find that it lies at a position slightly cooler and more luminous than the cluster's main sequence. This is consistent with our conclusion that the star has a slightly higher age and/or lower helium content than the cluster. \mucas\ is possibly the oldest star in the sky visible to the naked eye. 

The dominant contributor to the error budget for the mass of \mucas~A is the uncertainty in the trigonometric parallax. It may be possible to reduce this uncertainty with {\it Gaia\/} observations, although the star was too bright to be included in DR2. The angular-diameter measurement could be improved using CHARA observations in the $V$ or $H$ bands. It would be important also to reconcile the various determinations of the effective temperature, and to redetermine the chemical composition based on consistent atmospheric parameters. 

This work was inspired by the classical paper of Dennis (1965). We are closer now to achieving his goal of a precise helium content for this ancient metal-poor star, with its cosmological implications, but further work remains to be done.

\acknowledgments

We are indebted to the numerous observers who painstakingly accumulated data on this difficult and important star over several decades. 
Support was provided by NASA through grants from the Space Telescope Science Institute, which is operated by the Association of Universities for Research in Astronomy, Inc., under NASA contract NAS5-26555. G.H.S. acknowledges support from NSF grant AST-1636624. We thank the referee for urging us to consider the effects of CTI in the WFPC2 images.
Useful comments and unpublished data were provided by J.~Drummond, G.~Gatewood, L.~Roberts, and J.~Russell.
Randal Telfer provided important information related to chromatic aberration in the WFC3 camera. 
Advice and support during the planning, execution, and analysis of the \HST\/ observations were provided by STScI Contact Scientists, Program Coordinators, and staff members:
S.~Baggett,
J.~Biretta,
G.~Hartig,
R.~Lucas,
J.~Mack,
T.~Royle,
E.~Sabbi,
K.~Sahu,
D.~Taylor,
and A.~Vick.
We thank Bengt Gustafsson for helpful comments on chemical abundance
determinations.

\facility{CHARA, {\it Gaia}, {\it Hipparcos}, \HST\/ (WFPC2, WFC3)} 

\clearpage

\appendix

\section{Correcting for Charge-Transfer Inefficiency in \HST\/ WFPC2 F953N Images
\label{sec:appendixa}}

As described in \S\ref{sec:wfpc2f953n}, we found evidence for CTI-induced positional offsets in our WFPC2 astrometry, at a level reaching about 5~mas toward the end of WFPC2's lifetime. To compensate for this effect, we added a final analysis step for the WFPC2 measurements, involving an empirical characterization of possible CTI effects. We describe these corrections in this Appendix.

The WFPC2 observations covered 10.3 years, reaching nearly
the end of the instrument lifetime, when CTI losses had become largest.
CTI is expected to grow roughly linearly in time, and primarily affects
inferred positions of faint objects through distortion of the PSF in the 
detector's $y$-direction, which is the direction of charge readout.  Our WFPC2 observations happened to be well posed for empirically 
determining astrometric CTI effects, with many observations being taken about six months
apart, at telescope orientations differing by about $180^\circ$.  

We derived right-ascension and declination residuals from an initial orbital fit to the motion of \mucas~B around A\null. These showed a signature of alternating residuals, especially in right ascension, that were correlated with telescope roll angle, and growing with time. We transformed the residuals into detector $x,y$ deviations. These deviations were then fit to linear functions of time.  The fit to the $x$ deviations was not statistically
significant, with a 30\% chance that random variations could explain the linear
correlation.  In $y$, however, where CTI is expected to influence position determinations,
there was only a 0.8\% chance that the derived linear correlation could result
from random errors. A least-squares fit yielded a progressive shift in $y$ of 0.12~pixel over the 10.3-year observing baseline (equivalent to a shift in $y$ reaching $0\farcs0054$ at the end of the 10.3-year interval). We applied these corrections to the astrometry, and then transformed the $x,y$ positions back to right ascension and declination, and then to position angle and separation. The CTI-adjusted WFPC2 astrometry is listed in the first 20 lines in Table~\ref{table:hstastrometry}. Application of this CTI correction eliminated the previously noted alternating deviations in right ascension, and dropped the 
residual scatter by about a factor of two.

For reference, Table~\ref{table:f953ndata} lists the WFPC2 astrometry we obtained before making the CTI corrections.

\begin{deluxetable}{lccc}[!hb]
\tablewidth{0 pt}
\tabletypesize{\footnotesize}
\tablecaption{Uncorrected WFPC2/F953N Astrometric Measurements of \mucas~B Relative to \mucas~A
\label{table:f953ndata}
}
\tablehead{
\colhead{UT Date} &
\colhead{Besselian} &
\colhead{Separation} &
\colhead{J2000 Position} \\
\colhead{} &
\colhead{Date} &
\colhead{[arcsec]} &
\colhead{Angle [$^\circ$]} 
}
\startdata
\noalign{\smallskip}
1997 Jul 04 & 1997.5057 & $0.4189\pm0.0010$ & $226.391\pm0.092$ \\  
1998 Jan 02 & 1998.0047 & $0.4454\pm0.0009$ & $214.246\pm0.066$ \\ 
1998 Jul 22 & 1998.5544 & $0.4077\pm0.0002$ & $200.461\pm0.075$ \\ 
1999 Feb 28 & 1999.1612 & $0.3490\pm0.0004$ & $179.567\pm0.087$ \\ 
1999 Aug 04 & 1999.5906 & $0.3150\pm0.0007$ & $160.830\pm0.101$ \\ 
2000 Feb 01 & 2000.0868 & $0.3226\pm0.0006$ & $136.991\pm0.071$ \\ 
2000 Jul 15 & 2000.5390 & $0.3624\pm0.0005$ & $118.189\pm0.064$ \\ 
2001 Jan 15 & 2001.0406 & $0.4324\pm0.0004$ & $102.605\pm0.067$ \\ 
2001 Jul 30 & 2001.5773 & $0.5224\pm0.0003$ & $ 91.290\pm0.057$ \\ 
2002 Jan 17 & 2002.0476 & $0.6119\pm0.0004$ & $ 84.426\pm0.040$ \\ 
2002 Aug 05 & 2002.5934 & $0.7085\pm0.0003$ & $ 78.491\pm0.059$ \\ 
2003 Feb 11 & 2003.1144 & $0.8049\pm0.0003$ & $ 74.159\pm0.036$ \\ 
2003 Aug 05 & 2003.5942 & $0.8838\pm0.0006$ & $ 70.919\pm0.036$ \\ 
2004 Jan 29 & 2004.0772 & $0.9662\pm0.0002$ & $ 68.329\pm0.037$ \\ 
2004 Aug 08 & 2004.6039 & $1.0403\pm0.0006$ & $ 65.848\pm0.040$ \\ 
2005 Jan 15 & 2005.0412 & $1.1087\pm0.0003$ & $ 64.165\pm0.035$ \\ 
2005 Aug 13 & 2005.6175 & $1.1796\pm0.0005$ & $ 61.853\pm0.035$ \\ 
2006 Jan 30 & 2006.0815 & $1.2400\pm0.0003$ & $ 60.490\pm0.035$ \\ 
2006 Sep 26 & 2006.7400 & $1.3014\pm0.0004$ & $ 58.610\pm0.034$ \\ 
2007 Oct 17 & 2007.7933 & $1.3887\pm0.0003$ & $ 55.856\pm0.032$ \\ 
\enddata
\end{deluxetable}


\section{Correcting for Chromatic Aberration in \HST\/ WFC3 F225W Images
\label{sec:appendixb}}

As discussed in \S\ref{sec:wfc3f225w}, we found it necessary to make corrections to our astrometric measurements of the \mucas\ binary that were made on frames obtained in the \HST\/ WFC3 F225W bandpass. Although the WFC3 camera primarily uses reflective optics, the color filters and front windows of the CCD module are transmissive elements. Thus there is a small amount of chromatic aberration in the camera. To compensate for this, the thicknesses of the filters were adjusted to maintain confocality and a common plate scale, at the nominal wavelength of each filter. This approach becomes an important problem in the case of \mucas,  because there is a significant red leak in the F225W filter, combined with the fact that \mucas~B is an M~dwarf, which is considerably redder than the G-type primary star.

To assess the size and direction of this effect in the \HST\/ images, we first obtained the system throughput curve of the camera plus F225W filter from the WFC3 website\footnote{\url{ http://www.stsci.edu/hst/instrumentation/wfc3/performance/throughputs}} at the Space Telescope Science Institute (STScI)\null. We convolved this with the spectral-energy distributions for stars of spectral types G8~V and M4~V, taken from the observational stellar-flux library assembled by \citet{Pickles1998}; these are also conveniently available from STScI.\footnote{\url{https://ssb.stsci.edu/cdbs_open/cdbs/deliveries/etc/trds.24.3xxxx/grid/pickles/dat_uvk/}} This convolution shows that the flux from \mucas~B is almost entirely detected through the red leak of the F225W filter. We find an effective wavelength for the B component of about 8600~\AA\null. Even the detected light from \mucas~A is mostly at the long-wavelength side of the filter's main bandpass; we find an effective wavelength for the A component of about 2500~\AA\null. 

We are grateful to Randal Telfer, Astronomical Optics Scientist at STScI, for providing us with information on the differential chromatic aberration in WFC3 images, from which we calculated approximate corrections to be applied to our astrometry. The dispersion induced by the windows and filter has two effects: (1)~an overall shift in image position across the field due primarily to a small tilt in the detector windows, and (2)~an increase in magnification at longer wavelengths. The size of these effects is predictable, using the known wavelength dependence of the index of refraction of fused silica.

For the image offset, Telfer provided a tabulation for the F225W filter as a function of wavelength. The offsets in detector $x,y$ coordinates increase from zero at the filter's central wavelength, to $(\Delta x, \Delta y) = (-0.008, +0.091)$~pixel at 2500~\AA, and $(-0.037, +0.407)$~pixel at 9000~\AA\null. By interpolating in the table to the effective wavelengths for \mucas~A and B, we find a net offset of the B image relative to A of $(\Delta x, \Delta y) = (-0.025, +0.271)$~pixel.

The effect of the higher magnification at longer wavelengths is to move the image of B in the direction away from the center of the detector relative to the position of the bluer A component. By interpolation in the tabulation of magnification versus wavelength, we find a difference in magnification between that for B and A of 0.00052.

Combining these, the differential offset of B relative to A in pixels becomes
$$\Delta x = -0.025 + 0.00052\, (x-x_0) \, ,$$
$$\Delta y = +0.271 + 0.00052\, (y-y_0) \, ,$$
where $(x,y)$ is the pixel position of the star, and $(x_0,y_0)$ is the pixel position of the center of the detector, nominally (2048, 2048) pixels.

For the F225W observation in 2010 January, we used the UVIS1-M512-SUB subarray, placing the stars at nominal position $(x,y) = (2048, 2304)$, and resulting in an offset of \mucas~B relative to A of $(\Delta x, \Delta y) = (-0.025, +0.404)$ pixel. The rest of our F225W data used the UVIS2-C512C-SUB subarray, nominally centered at $(x,y) = (256, 256)$. At this location, the relative offset is $(\Delta x, \Delta y) = (-0.957, -0.661)$~pixel. In arcseconds the offsets are small: $(-0\farcs001, +0\farcs016)$ for the first 2010 observation, and $(-0\farcs038, -0\farcs026)$ for the rest; but they are large compared to the precision of the \HST\/ astrometry.

Table~\ref{table:f225wdata} lists the astrometric measurements of \mucas~B before applying the above corrections. The $(x,y)$ positions on the detector have been converted to separation and PA on the sky, using the known plate scale and orientation of the spacecraft, as described in \S\ref{sec:analysis}. These  values, adjusted for chromatic aberration, are included in the main text, in Table~\ref{table:hstastrometry}.

\begin{deluxetable}{lccc}[!]
\tablewidth{0 pt}
\tabletypesize{\footnotesize}
\tablecaption{Uncorrected WFC3/F225W Astrometric Measurements of \mucas~B Relative to \mucas~A
\label{table:f225wdata}
}
\tablehead{
\colhead{UT Date} &
\colhead{Besselian} &
\colhead{Separation} &
\colhead{J2000 Position} \\
\colhead{} &
\colhead{Date} &
\colhead{[arcsec]} &
\colhead{Angle [$^\circ$]} 
}
\startdata
\noalign{\smallskip}
2010 Jan 09 & 2010.0236 & $1.4730\pm0.0037$ & $ 50.142\pm0.073$ \\ 
2010 Dec 03 & 2010.9222 & $1.4926\pm0.0056$ & $ 50.046\pm0.283$ \\ 
2011 Dec 05 & 2011.9264 & $1.4464\pm0.0017$ & $ 47.613\pm0.075$ \\ 
2012 Dec 02 & 2012.9204 & $1.3590\pm0.0010$ & $ 45.046\pm0.098$ \\ 
2013 Oct 25 & 2013.8179 & $1.2262\pm0.0007$ & $ 43.050\pm0.125$ \\ 
2015 Jan 06 & 2015.0150 & $1.0372\pm0.0007$ & $ 38.375\pm0.080$ \\ 
\enddata
\end{deluxetable}


\clearpage 

\begin{thebibliography}{}

\bibitem[Abt et al.(1980)]{Abt1980} Abt, H.~A., Sanwal, N.~B., \& Levy, S.~G.\ 1980, \apjs, 43, 549

\bibitem[Abt \& Willmarth(1987)]{Abt1987} Abt, H.~A., \& Willmarth, D.~W.\ 1987, \apj, 318, 786

\bibitem[Abt \& Willmarth(2006)]{Abt2006} Abt, H.~A., \& Willmarth, D.\ 2006, \apjs, 162, 207

\bibitem[Adams \& Joy(1919)]{Adams1919} Adams, W.~S., \& Joy, A.~H.\
1919, \apj, 49, 179

\bibitem[Agati et al.(2015)]{Agati2015} Agati, J.-L., Bonneau, D.,
Jorissen, A., et al.\ 2015, \aap, 574, A6 

\bibitem[Bach(2015)]{Bach2015} Bach, K.\ 2015, Journal of Korean Astronomical Society, 48, 165

\bibitem[Beavers \& Eitter(1986)]{Beavers1986} Beavers, W.~I., \& Eitter, J.~J.\ 1986, \apjs, 62, 147

\bibitem[Bennett et al.(2013)]{blw13}
Bennett, C.~L., Larson, D., Weiland, J.~L., et al.~2011, ApJS, 208, 20

\bibitem[Bergbusch \& Stetson(2009)]{bs09}
Bergbusch, P.~A., \& Stetson, P.~B.~2009, AJ, 138, 1455

\bibitem[Bertelli Motta et al.(2018)]{bpr18}
Bertelli Motta, C., Pasquali, A., Richer, J., et al.~2018, MNRAS, 478, 425

\bibitem[Bond et al.(2015)]{Bond2015} Bond, H.~E., Gilliland, R.~L., Schaefer, G.~H., et al.\ 2015, \apj, 813, 106 (B15)

\bibitem[Bond et al.(2018)]{Bond2018} Bond, H.~E., Gilliland, R.~L., Schaefer, G.~H., et al.\ 2018, Research Notes of the American Astronomical Society, 2, 147

\bibitem[Bond et al.(2017)]{Bond2017} Bond, H.~E., Schaefer, G.~H., Gilliland, R.~L., et al.\ 2017, \apj, 840, 70 (B17)

\bibitem[Boyajian et al.(2008)]{Boyajian2008} Boyajian, T.~S., McAlister, H.~A., Baines, E.~K., et al.\ 2008, \apj, 683, 424 (B08)

\bibitem[Brogaard et al.(2017)]{bvb17}
Brogaard, K., VandenBerg, D.~A., Bedin, L.~R., Milone, A.~P., Thygesen, A., \&
Grundahl, F.~2017, MNRAS, 468, 645

\bibitem[Campbell(1901)]{Campbell1901} Campbell, W.~W.\ 1901, \apj, 13, 98 

\bibitem[Capitanio et al.(2017)]{clv17}
Capitanio, L., Lallement, R., Vergely, J.~L., Elyajouri, M., \& Monreal-Ibero,
A.~2017, A\&A, 606, A65

\bibitem[Carretta et al.(2009)]{cbg09}
Carretta, E., Bragaglia, A., Gratton, R.~G., D'Orazi, V., \& Lucatello,
 S.~2009, A\&A, 508, 695

\bibitem[Casagrande et al.(2014)]{Casagrande2014} Casagrande, L., Portinari, L., Glass, I.~S., et al.\ 2014, \mnras, 439, 2060

\bibitem[Casagrande et al.(2010)]{crm10}
Casagrande, L., Ram\'irez, I., Mel\'endez, J., Bessell, M., \& Asplund,
M.~2010, A\&A, 512, 54

\bibitem[Casagrande \& VandenBerg(2014)]{cv14}
Casagrande, L., \& VandenBerg, D.~A.~2014, MNRAS, 444, 392\ \ \ (CV14)

\bibitem[Casamiquela et al.(2020)]{Casamiquela2020} Casamiquela, L., Tarricq, Y., Soubiran, C., et al.\ 2020, \aap, 635, A8


\bibitem[Chen et al.(2018)]{crc18}
Chen, S., Richer, H., Caiazzo, I., \& Heyl, J.~2018, ApJ, 867, 132

\bibitem[Cordero et al.(2014)]{cpj14} Cordero, M.~J., Pilachowski, C.~A., Johnson, C.~I., et al.\ 2014, \apj, 780, 94

\bibitem[Cyburt et al.(2016)]{cfo16}
Cyburt, R.~H., Fields, B.~D., Olive, K.~A., \& Tsung-Han, Y.~2016, RvMP,
88, 015004

\bibitem[Denissenkov et al.(2017)]{dvk17}
Denissenkov, P.~A., VandenBerg, D.~A., Kopacki, G., \& Ferguson, J.~W.~2017,
ApJ, 849, 159

\bibitem[Dennis(1965)]{Dennis1965} Dennis, T.~R.\ 1965, \pasp, 77, 283
(D65)

\bibitem[Drummond et al.(1995)]{Drummond1995} Drummond, J.~D., Christou, J.~C., \& Fugate, R.~Q.\ 1995, \apj, 450, 380

\bibitem[Duquennoy et al.(1991)]{Duquennoy1991} Duquennoy, A., Mayor, M., \& Halbwachs, J.-L.\ 1991, \aaps, 88, 281

\bibitem[Faulkner(1971)]{Faulkner1971} Faulkner, J.\ 1971, \prl, 27, 206

\bibitem[Feibelman(1976)]{Feibelman1976} Feibelman, W.~A.\ 1976, \apj, 209, 497

\bibitem[Gaia Collaboration et al.(2018)]{Gaia2018} Gaia
Collaboration, Brown, A.~G.~A., Vallenari, A., et al.\ 2018, \aap, 616, A1 

\bibitem[Gilliland(2005)]{Gilliland2005} Gilliland, R.~L.\ 2005, Telescope Instrument Science Report 2005-02 (Baltimore,
MD: STScI)

\bibitem[Gruyters et al.(2014)]{gnk14} Gruyters, P., Nordlander, T., \& Korn, A.~J.~2014, A\&A, 567, A72

\bibitem[Gustafsson et al.(2008)]{gee08}
Gustafsson, B., Edvardsson, B., Eriksson, K., Jorgensen, U.~G., Nordlund, 
\AA, \& Plez, B.~2008, A\&A, 486, 951

\bibitem[Harrington et al.(1993)]{Harrington1993} Harrington, R.~S., Dahn, C.~C., Kallarakal, V.~V., et al.\ 1993, \aj, 105, 1571

\bibitem[Haywood et al.(1992)]{Haywood1992} Haywood, J.~W., Hegyi, D.~J., \& Gudehus, D.~H.\ 1992, \apj, 392, 172

\bibitem[Heintz(1994)]{Heintz1994} Heintz, W.~D.\ 1994, \aj, 108, 2338

\bibitem[Heintz \& Cantor(1994)]{HeintzCantor1994} Heintz, W.~D., \& Cantor, B.~A.\ 1994, \pasp, 106, 363

\bibitem[Heiter et al.(2015)]{Heiter2015} Heiter, U., Jofr{\'e}, P., Gustafsson, B., et al.\ 2015, \aap, 582, A49

\bibitem[Holman \& Wiegert(1999)]{Holman1999} Holman, M.~J., \& Wiegert, P.~A.\ 1999, \aj, 117, 621

\bibitem[Horch et al.(2019)]{Horch2019} Horch, E.~P., Tokovinin, A., Weiss, S.~A., et al.\ 2019, \aj, 157, 56

\bibitem[Horch et al.(2015)]{Horch2015} Horch, E.~P., van Belle, G.~T., Davidson, J.~W., et al.\ 2015, \aj, 150, 151

\bibitem[Jao et al.(2016)]{Jao2016} Jao, W.-C., Nelan, E.~P., Henry, T.~J., et al.\ 2016, \aj, 152, 153

\bibitem[Jasniewicz \& Mayor(1988)]{Jasnie1988} Jasniewicz, G., \& Mayor, M.\ 1988, \aap, 203, 329

\bibitem[Jofr{\'e} et al.(2014)]{Jofre2014} Jofr{\'e}, P., Heiter, U., Soubiran, C., et al.\ 2014, \aap, 564, A133



\bibitem[Jofr{\'e} et al.(2018)]{Jofre2018} Jofr{\'e}, P., Heiter, U., Tucci Maia, M., et al.\ 2018, RNAAS, 2, 152

\bibitem[Johnson \& Morgan(1953)]{Johnson1953} Johnson, H.~L., \& Morgan, W.~W.\ 1953, \apj, 117, 313 

\bibitem[Karovicova et al.(2018)]{Karovicova2018} Karovicova, I., White, T.~R., Nordlander, T., et al.\ 2018, \mnras, 475, L81

\bibitem[Keenan \& Keller(1953)]{Keenan1953} Keenan, P.~C., \& Keller, G.\ 1953, \apj, 117, 241

\bibitem[Korn et al.(2007)]{kgr07}
Korn, A.~J., Grundahl, F., Richard, O., et al.~2007, ApJ, 671, 402

 
\bibitem[Lebreton et al.(1999)]{Lebreton1999} Lebreton, Y., Perrin, M.-N., Cayrel, R., et al.\ 1999, \aap, 350, 587
 
\bibitem[Lind et al.(2012)]{lba12}
Lind, K., Bergemann, M., \& Asplund M.~2012, MNRAS, 427, 50

\bibitem[Lindegren et al.(2018)]{lhb18}
Lindegren, L., Hernandez, J., Bombrun, A., et al.~2018, A\&A, 616. A2


\bibitem[Lippincott(1981)]{Lippincott1981} Lippincott, S.~L.\ 1981, \apj, 248, 1053 

\bibitem[Lippincott \& Wyckoff(1964)]{Lippincott1964} Lippincott, S.~L., \& Wyckoff, S.\ 1964, \aj, 69, 471

\bibitem[Luck(2017)]{Luck2017} Luck, R.~E.\ 2017, \aj, 153, 21

\bibitem[Luck \& Heiter(2005)]{Luck2005} Luck, R.~E., \& Heiter, U.\ 2005, \aj, 129, 1063 

\bibitem[Mamajek \& Hillenbrand(2008)]{mh08}
Mamajek, E.~E., \& Hillenbrand, L.~A.~2008, ApJ, 687, 1264

\bibitem[Marino et al.(2016)]{mmc16}
Marino, A.~F., Milone, A.~P., Casagrande, L., et al.~2016, MNRAS, 459, 610

\bibitem[Mermilliod(1991)]{me91}
Mermilliod, J.~C.~1991, Catalogue of Homogeneous Means in the $UBV$ System
 (Lausanne: Univ.~de Lausanne)

\bibitem[Miczaika(1940)]{Miczaika1940} Miczaika, G.\ 1940, Astronomische Nachrichten, 270, 249

\bibitem[Nordlander et al.(2012)]{nkr12}
Nordlander, T., Korn, A.~J., Richard, O., \& Lind, K.~2012, ApJ, 753, 48

\bibitem[Nordstr\"om et al.(2004)]{nma04}
Nordstr\"om, B., Mayor, M., Andersen, J., et al.~2004, A\&A, 418, 989

\bibitem[{\"O}nehag et al.(2014)]{ogk14} {\"O}nehag, A., Gustafsson, B., \& Korn, A.\ 2014, \aap, 562, A102

\bibitem[Oort(1926)]{Oort1926} Oort, J.~H.\ 1926, Ph.D. Thesis, Leiden University

\bibitem[Pickles(1998)]{Pickles1998} Pickles, A.~J.\ 1998, \pasp, 110, 863

\bibitem[Pierce \& Lavery(1985)]{Pierce1985} Pierce, M.~J., \& Lavery, R.~J.\ 1985, \aj, 90, 647


\bibitem[Planck Collaboration et al.(2014)]{pla14} Planck Collaboration, Ade, P.~A.~R., Aghanim, N., et al.\ 2014, \aap, 571, A16

\bibitem[Richard et al.(2002)]{rmr02} Richard, O., Michaud, G., Richer, J., et al.\ 2002, \apj, 568, 979

\bibitem[Roman(1955)]{Roman1955} Roman, N.~G.\ 1955, \apjs, 2, 195

\bibitem[Russell \& Gatewood(1984)]{Russell1984} Russell, J.~L., \& Gatewood, G.~D.\ 1984, \pasp, 96, 429


\bibitem[Schlafly \& Finkbeiner(2011)]{sf11}
Schlafly, E.~F., \& Finkbeiner, D.~P.~2011, ApJ, 737, 103

\bibitem[Schlegel et al.(1998)]{sfd98}
Schlegel, D., Finkbeiner, D.~P., \& Davis, M.~1998, ApJ, 500, 525

Soubiran, C., \& Girard, P.~2005, A\&A, 438, 139

\bibitem[Soubiran et al.(2016)]{Soubiran2016} Soubiran, C., Le Campion, J.-F., Brouillet, N., et al.\ 2016, \aap, 591, A118

\bibitem[Spite \& Spite(1982)]{ss82} Spite, F., \& Spite, M.~1982, A\&A, 115, 357

\bibitem[Thompson et al.(2020)]{Thompson2020} Thompson, I.~B., Udalski, A., Dotter, A., et al.\ 2020, \mnras, 492, 4254

\bibitem[Thygesen et al.(2018)]{tsa18}
Thygesen, A.~O., Sbordone, L., Andrievsky, A., et al.~2018, A\&A, 572, A108

\bibitem[VandenBerg et al.(2014)]{vbf14}
VandenBerg, D.~A., Bergbusch, P.~A., Ferguson, J.~W., \& Edvardsson, B.~2014, 
ApJ, 794, 72

\bibitem[VandenBerg et al.(2013)]{vbl13}
VandenBerg, D.~A., Brogaard, K., Leaman, R., \& Casagrande, L.~2013, ApJ, 775,
 134\ \ \ (VBLC13)

\bibitem[VandenBerg et al.(2010)]{vcs10}
VandenBerg, D.~A., Casagrande, L., \& Stetson, P.~B.~2010, ApJ, 140, 1020

\bibitem[VandenBerg et al.(2002)]{vrm02} VandenBerg, D.~A., Richard, O., Michaud, G., et al.\ 2002, \apj, 571, 487

\bibitem[van den Bos(1964)]{vandenBos1964} van den Bos, W.~H.\ 1964, in Astronomical Techniques, ed.\ W. A. Hiltner (Chicago: Univ.\ Chicago Press), 537

\bibitem[van Leeuwen(2007)]{vanLeeuwen2007} van Leeuwen, F.\ 2007, \aap, 474, 653

\bibitem[Wagman(1961)]{Wagman1961} Wagman, N. E. 1961, \aj, 66, 433

\bibitem[Wagman et al.(1963)]{Wagman1963} Wagman, N.~E., Daniel, Z., \&
Crissman, G.~B.\ 1963, \aj, 68, 352 

\bibitem[Wang et al.(2017)]{wpc17} Wang, Y., Primas, F., Charbonnel, C., et al.\ 2017, \aap, 607, A135

\bibitem[Wehinger \& Wyckoff(1966)]{Wehinger1966} Wehinger, P.~A., \&
Wyckoff, S.\ 1966, \aj, 71, 185 

\bibitem[White et al.(2018)]{White2018} White, T.~R., Huber, D., Mann, A. W., et al.\ 2018, \mnras, 477, 4403

\bibitem[Wickes(1975)]{Wickes1975} Wickes, W.~C.\ 1975, \aj, 80, 655

\bibitem[Wickes \& Dicke(1974)]{Wickes1974} Wickes, W.~C., \& Dicke,
R.~H.\ 1974, \aj, 79, 1433 

\bibitem[Worek \& Beardsley(1977)]{Worek1977} Worek, T.~F., \& Beardsley, W.~R.\ 1977, \apj, 217, 134


\end{thebibliography}
\end{document}